\documentclass[journal, 12pt, onecolumn, draftclsnofoot]{IEEEtran}

\usepackage[reqno]{amsmath}
\usepackage{graphicx}
\usepackage{bbm}
\usepackage{mathrsfs}
\usepackage{stmaryrd}
\usepackage{graphics}
\usepackage{acronym}
\usepackage{longtable}
\usepackage{mathtools}
\usepackage{times}
\usepackage{setspace}
\usepackage{cite}
\usepackage{array}
\usepackage{subfigure}
\usepackage{amsmath,amsthm}
\usepackage{amssymb}
\usepackage{wasysym,url}
\usepackage{setspace,float}
\usepackage{color}
\usepackage{cases,bm}
\usepackage[inline]{enumitem}
\usepackage{tikz}
\usepackage{hyperref}
\usepackage{mathtools,cuted}
\usepackage{multirow}
\usepackage{array}
\usepackage{booktabs}
\usepackage[noabbrev]{cleveref}
\crefformat{figures}{(#2#1#3)}
\usepackage{xcolor}
\usepackage{tabularx}
\usepackage{booktabs}
\usepackage[linesnumbered,ruled,vlined]{algorithm2e}
\DontPrintSemicolon

\makeatletter
\newcommand{\leqnomode}{\tagsleft@true\let\veqno\@@leqno}
\newcommand{\reqnomode}{\tagsleft@false\let\veqno\@@eqno}
\makeatother

\newcommand{\norm}[1]{\left\Vert{#1}\right\Vert}
\newcommand{\abs}[1]{\left\vert{#1}\right\vert}
\newcommand{\bb}[1]{\mathbb{#1}}

\newcommand{\bd}[1]{\mathbf{#1}}
\newcommand{\bld}[1]{\boldsymbol{#1}}

\newcommand{\cl}[1]{\mathcal{#1}}
\newcommand{\fr}[1]{\mathfrak{#1}}

\newcommand{\sT}{\mathscr{T}}
\newcommand{\sE}{\mathscr{E}}

\newcommand{\TT}{\mathsf{T}}
\newcommand{\jj}{\mathrm{j}}
\newcommand{\dd}{\mathrm{d}}

\newcommand{\sinc}{\mathrm{sinc}}

\newcommand{\orcid}[1]{\href{https://orcid.org/#1}{\textcolor[HTML]{A6CE39}{\aiOrcid}}}

\newtheorem{theorem}{Theorem}
\newtheorem{lemma}{Lemma}

\newtheorem{corollary}{Corollary}

\newtheorem{definition}{Definition}

\definecolor{lime}{HTML}{A6CE39}
\DeclareRobustCommand{\orcidicon}{%
	\begin{tikzpicture}
	\draw[lime, fill=lime] (0,0)
	circle [radius=0.16]
	node[white] {{\fontfamily{qag}\selectfont \tiny ID}};
	\draw[white, fill=white] (-0.0625,0.095)
	circle [radius=0.007];
	\end{tikzpicture}
	\hspace{-2mm}
}

\foreach \x in {A, ..., Z}{%
	\expandafter\xdef\csname orcid\x\endcsname{\noexpand\href{https://orcid.org/\csname orcidauthor\x\endcsname}{\noexpand\orcidicon}}
}

\title{Neuromorphic Sampling of Signals in Shift-Invariant Spaces}
\author{\IEEEauthorblockN{
Abijith~Jagannath~Kamath\orcidA{}~\IEEEmembership{Student~Member,~IEEE}, and~Chandra~Sekhar~Seelamantula\orcidC{}~\IEEEmembership{Senior~Member,~IEEE}
}

\thanks{
A.~J.~Kamath and C.~S.~Seelamantula are with the Department of Electrical Engineering, Indian Institute of Science, Bangalore (Email: \{abijithj, css\}@iisc.ac.in).\\
\indent The figures in this paper are in colour in the electronic version.
}}

\begin{document}
\maketitle

%


\begin{abstract}
Neuromorphic sampling is a paradigm shift in analog-to-digital conversion where the acquisition strategy is opportunistic and measurements are recorded only when there is a significant change in the signal. Neuromorphic sampling has given rise to a new class of \emph{event-based} sensors called \emph{dynamic vision sensors} or \emph{neuromorphic cameras}. The neuromorphic sampling mechanism utilizes low power and provides high-dynamic range sensing with low latency and high temporal resolution. The measurements are sparse and have low redundancy making it convenient for downstream tasks. In this paper, we present a sampling-theoretic perspective to neuromorphic sensing of continuous-time signals. We establish a close connection between neuromorphic sampling and \emph{time-based sampling} --- where signals are encoded temporally. We analyse neuromorphic sampling of signals in shift-invariant spaces, in particular, bandlimited signals and polynomial splines. We present an iterative technique for perfect reconstruction subject to the events satisfying a density criterion. We also provide necessary and sufficient conditions for perfect reconstruction. Owing to practical limitations in meeting the sufficient conditions for perfect reconstruction, we extend the analysis to approximate reconstruction from sparse events. In the latter setting, we pose signal reconstruction as a continuous-domain linear inverse problem whose solution can be obtained by solving an equivalent finite-dimensional convex optimization program using a variable-splitting approach. We demonstrate the performance of the proposed algorithm and validate our claims via experiments on synthetic signals.
\end{abstract}





\section{Introduction}
\begin{figure}[t]
	\centering
	\includegraphics[width=3.3in]{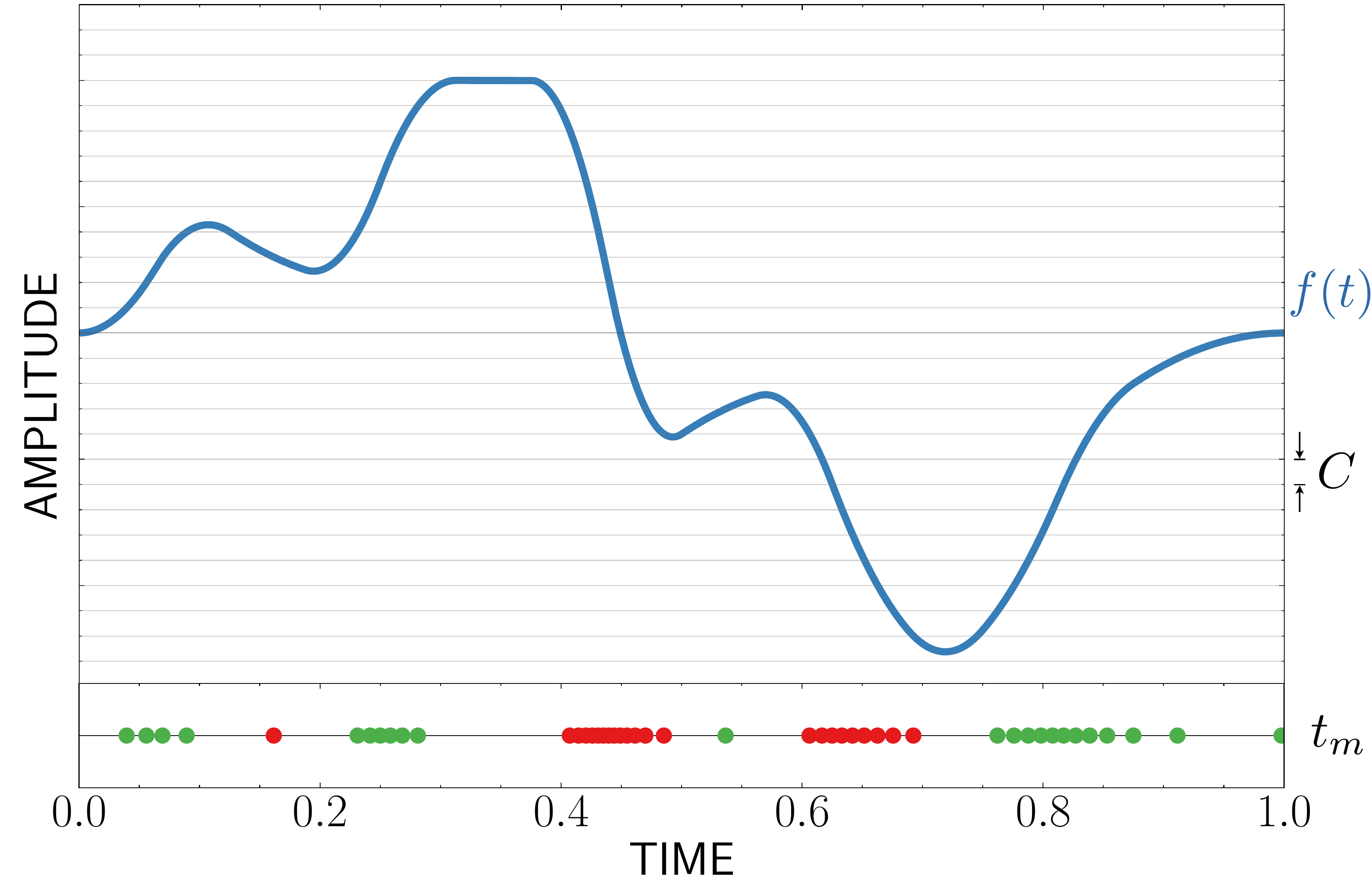}
	\caption{Illustration of neuromorphic sampling of a continuous-time signal $f(t)$ with the temporal contrast threshold $C$. The signal is encoded in terms of the time instants $t_m, \; m=1,2,\ldots$, and the corresponding sign of change, i.e., ON (green) or OFF (red). In intervals of no activity, the neuromorphic sampling scheme does not output events.}
	\label{fig:ttransform_demo}
\end{figure}

\begin{figure}[t!]
	\centering
	\includegraphics[width=3in]{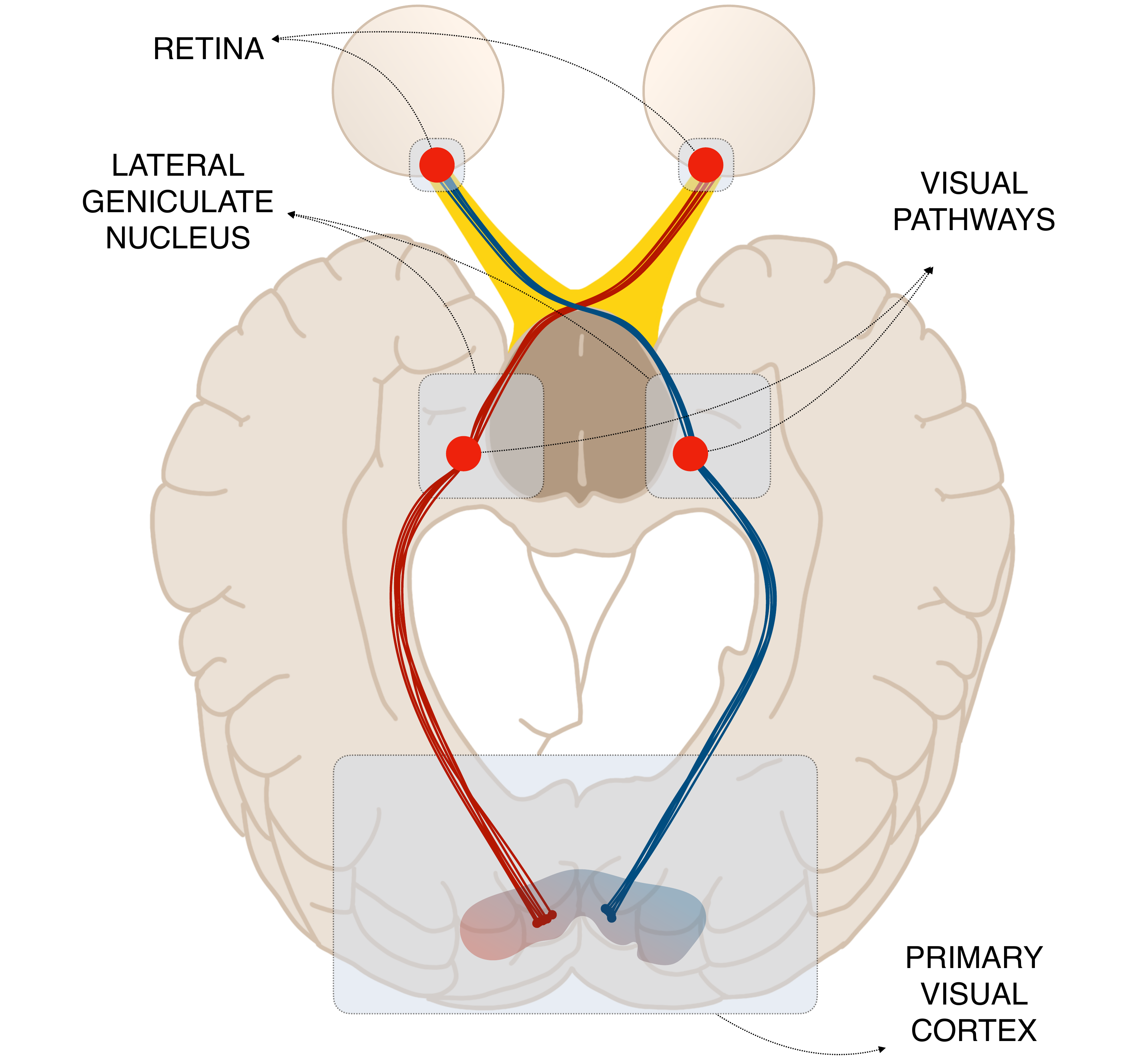}
	\caption{A cross-section of the human vision system: Visual stimulus incident on the retina is carried to the lateral geniculate nucleus, where the encoding occurs. The encoded information is carried to the primary visual cortex where the information is perceived via magnocellular and parvocellular pathways.}
	\label{fig:visual_pathways}
\end{figure}

\IEEEPARstart{N}{euromorphic} sampling is a novel analog-to-digital (A/D) acquisition scheme that is different from uniform sampling \cite{shannon1949communication}. In neuromorphic sampling, continuous-time signals are encoded using temporal measurements that are signal-dependent, unlike amplitude samples recorded at fixed, uniform intervals. The temporal measurements correspond to changes in the absolute value of the signal by a fixed constant, along with the corresponding polarity indicating the sign of the change. The temporal measurement and the polarity (ON/OFF) together constitute an {\it event}. Thence, neuromorphic sampling falls in the category of {\it event-driven sampling}, and is inherently opportunistic. Figure~\ref{fig:ttransform_demo} demonstrates neuromorphic sampling of a continuous-time signal $f(t)$. The event-time instants $\{t_m\}$ are typically nonuniform, and are {\it dense} in regions of significant change in the signal and sparse in regions where the change is small. Therefore, events are not recorded in regions of inactivity in the signal. This is in contrast to uniform sampling, where the measurements are always acquired at a fixed rate, and are signal-independent.\\
\indent Neuromorphic sampling has spurred the development of a new class of vision sensors called {\it dynamic vision sensors (DVS)}, also known as {\it event cameras} or {\it neuromorphic cameras} \cite{mahowald1994silicon,delbruck2010sensors,lichtsteiner2008sensor,brandii2014sensor}. Event cameras such as the DAVIS346 \cite{gallego2022event} are energy efficient ($\approx 1$ mW), and the pixels in the sensor array asynchronously measure events without relying on a global clock. They are designed to provide compressed measurements at the source and minimize redundancy in the representation. Further, since the sensing is relative to the previous measurement, high-dynamic-range (HDR) signals ($\approx 120$ dB) can be accommodated without the sensor getting saturated. Event cameras have a high temporal resolution ($\approx 1~\mu$s), which makes them perfectly suited for sensing ultrafast changes in the visual scene.\\
\indent Neuromorphic sensor development is inspired by the functioning of the human vision system \cite{van2003selective}. Figure~\ref{fig:visual_pathways} shows a schematic of the cross-section of the human vision system. The image formed on the retina is transmitted to the lateral geniculate nucleus (LGN) by \emph{nasal} and \emph{temporal} connections, and thereafter to the primary visual cortex after encoding \cite{van2003selective}. The encoded information is carried via two types of visual pathways --- parvocellular and magnocellular. The parvocellular pathway is responsible for encoding a sustained visual stimulus such as the subject background and peripheral vision. It encodes low-contrast colour vision with a high spatial resolution. On the other hand, the magnocellular pathway is responsible for encoding transient vision, i.e., visual events that are fast-changing. It encodes high-contrast monochrome vision with a low spatial resolution. Neuromorphic cameras are inspired by the magnocellular sensing mechanism, whereas standard frame-based cameras mimic the parvocellular mechanism.


\subsection{Related Literature}
\label{sec:related_literature}
The techniques in the literature that come closest to neuromorphic sensing are those based on time-encoding machines (TEMs), which are also event-driven and have origins in how neural systems respond to stimuli. We review key developments in TEMs, which are required to appreciate neuromorphic sensing from the perspective of time-encoding.\\
\indent Lazar and T\'oth developed the {\it integrate-and-fire time-encoding machine} (IF-TEM) \cite{lazar2003recovery,lazar2004sensitivity}, motivated by the representation of neural signals as a sequence of action potentials \cite{adrian1928basis,burkitt2006review}. Whenever the accumulated stimulus exceeds a threshold, a spike is generated across the synapse \cite{adrian1928basis,burkitt2006review}. In the IF-TEM, a continuous-time signal is encoded using temporal measurements called {\it trigger times} that correspond to the event when the accumulated (or integrated) signal along with a bias exceeds a threshold. Lazar and Pnevmatikakis \cite{lazar2011video} used the IF-TEM for encoding videos using an array of spatiotemporal receptive fields. IF-TEMs have also given rise to a novel class of integrate-and-fire vision sensors called \emph{spiking cameras} \cite{zhu2022spiking}. Mart\'inez-Nuevo {\it et al.} \cite{martinez2019delta} introduced {\it amplitude sampling} using a delta-ramp encoder that uses a ramp signal of sufficient slope as bias to the input signal in order to make it monotonic such that the nonuniform temporal measurements are in one-to-one correspondence with the uniform samples of the biased signal.
\indent Lazar and T\'oth \cite{lazar2003recovery,lazar2004sensitivity} also addressed the problem of signal reconstruction from IF-TEM measurements. Their reconstruction technique is iterative and results in perfect reconstruction when the signals are bandlimited. Gontier and Vetterli \cite{gontier2014sampling} extended Lazar and T\'oth's analysis to achieve perfect reconstruction of signals belonging in shift-invariant spaces. Gontier and Vetterli's reconstruction technique is also iterative (projections onto convex sets (POCS) \cite{karen2020multichannel}) and effectively corresponds to solving a continuous-domain inverse problem \cite{sandberg1963properties,wiley1978recovery}. Alexandru and Dragotti \cite{alexandru2019reconstructing}, Rudresh {\it et al.} \cite{rudresh2020time,kamath2021time}, Naaman {\it et al.} \cite{naaman2022fritem} developed techniques for signal reconstruction from IF-TEM measurements for the analog-sparse case of signals with a finite rate of innovation. The techniques rely on high-resolution spectral estimation ideas, and perfect reconstruction is possible in the absence of noise.\\
\indent Gontier and Vetterli \cite{gontier2014sampling} also introduced the crossing time-encoding machine (C-TEM), which is a generalization of zero-crossing detectors. More precisely, the C-TEM measures the time instants where the input signal crosses a sinusoidal reference signal. Gontier and Vetterli showed that the C-TEM is closely linked to the IF-TEM, which enables the deployment of similar techniques for signal reconstruction.\\
\indent Kamath and Seelamantula \cite{kamath2022differentiate} introduced a new model, namely, the {\it differentiate-and-fire time-encoding machine} (DIF-TEM), which is complementary to the IF-TEM. In a DIF-TEM, the trigger times correspond to events when the sum of the derivative of the signal and a ramp exceeds a preset threshold. Signal reconstruction from DIF-TEM measurements is possible up to a constant offset. This is understandable given that the DIF-TEM sensing mechanism relies on changes only and a global constant offset would not make a difference to the event stream.\\
\indent Specifically, in the context of neuromorphic cameras, there have been several attempts at reconstructing images and videos from event cameras, starting from the work of Brandli {\it et al.} \cite{brandli2014real} who used a zero-order hold interpolation for reconstructing a temporally piecewise-constant video. Miyatani {\it et al.} \cite{barua2016direct} considered image intensity reconstruction using sparsity priors on a learnable dictionary. In a similar vein, Bardow {\it et al.} \cite{bardow2016simultaneous} and Reinbacher {\it et al.} \cite{reinbacher2016real} considered reconstruction from events as a regularized inverse problem. Several deep-learning based techniques have also been developed to solve the reconstruction problem \cite{wang2019event,rebecq2019events,rebecq2019high,wang2020event,scheerlinck2020fast,cadena2021spade,zhang2020video,liu2022convolutional}. Scheerlinck \cite{scheerlinck18accv} and Tulyakov \cite{tulyakov2021time} utilized the low-rate image frames acquired in addition to event streams in some neuromorphic cameras, and solved video reconstruction using interpolation techniques. {\it Binning} or temporal aggregation of events is a typical strategy to increase the density of measurements, although at the expense of loss of temporal resolution. In contrast with this line of work, we take a sampling-theoretic perspective to neuromorphic sensing. We establish a concrete link between neuromorphic sensing and time-encoding, which sets a solid foundation on which the problem of signal reconstruction from events can be addressed. To the best of our knowledge, this is the first such attempt at addressing the reconstruction problem within the framework of nonuniform sampling.


\subsection{Contributions of this Paper}
We seek to develop an accurate mathematical model for event generation and neuromorphic sampling of continuous-time signals. The objective is to derive necessary and sufficient conditions for perfect signal reconstruction starting from the event stream. To begin with, we present the necessary mathematical preliminaries related to shift-invariant spaces and time-encoding machines (Section~\ref{sec:prelims}). We argue that a neuromorphic sensor functions akin to a time-encoding machine. We achieve this by proposing an accurate event generation model for neuromorphic sensing and by specifying the corresponding \emph{$t$-transform} (Section~\ref{sec:event_generation}). We consider neuromorphic sampling of signals in shift-invariant spaces, in particular, signals in the Paley-Wiener space of finite-energy, bandlimited functions defined on the real line, and polynomial spline spaces. We show that, perfect reconstruction is possible using an alternating projections technique, and derive necessary and sufficient conditions (Section~\ref{sec:perfect_reconstruction}). When the conditions are met, the problem effectively boils down to solving an overdetermined system. In practical scenarios, where the conditions may not always be met, one would like to develop a computationally efficient, approximate reconstruction strategy. Toward this end, we show, via a continuous-domain formulation, that the problem of signal reconstruction from the event stream can be interpreted as solving a linear inverse problem subject to the measurement constraint. Since the neuromorphic sensing mechanisms typically gives rise to sparse measurements, leveraging sparsity, we develop an efficient, approximate signal reconstruction algorithm by using an $L_p$-(pseudo)norm minimization approach with $p>0$, subject to the measurement constraint (Section~\ref{sec:sparse_reconstruction}). We illustrate the reconstruction performance of both perfect reconstruction and approximate reconstruction techniques.


\section{Mathematical Preliminaries}
\label{sec:prelims}
In this section, we introduce the mathematical tools that are necessary for the remainder of the paper. The set of contiguous integers $\{a, a+1, \cdots, b\}$ is denoted as $\llbracket a,b\rrbracket$. Sequences defined on integers are denoted using bold symbols $\bld a=\{a_k\}_{k\in\bb Z}$. A sequence of iterates is denoted as $\{\bd a_k\}_{k=0}^\infty$, where $k$ is the iteration index. The class of $k$-times continuously differentiable functions defined on the real line $\bb R$ is denoted as $\cl C^k(\bb R)$. The $\ell_p$ and $L_p$ norms/pseudonorms are denoted by $\|\cdot\|_{\ell_p}$ and $\|\cdot\|_{L_p}$, respectively.


\subsection{Polynomial B-Splines}
\label{subsec:splines}
The causal polynomial B-spline \cite{unser1999splines} of degree zero denoted by $\beta_0$ is defined as
\begin{equation*}
	\beta_0(t) = \begin{cases}
	1, & 0 \leq t < 1, \\
	0, & \text{elsewhere}.
	\end{cases}
\end{equation*}
The causal B-spline $\beta_n$ of degree $n\in\bb N$ is given by the $n$-fold convolution of $\beta_0$:
\[
	\beta_n = \underbrace{\beta_0 * \beta_0 * \cdots * \beta_0}_{n\text{-fold}}.	
\]
Polynomial B-splines have a compact support. More precisely, the support of $\beta_n$ is $[0, n+1]$. A polynomial B-spline of degree $n-1$ is $(n+1)$-sparse, i.e., it has $(n+1)$ nonzero coefficients, under the $n^\text{th}$ order differential operator $\displaystyle\mathrm{D}^{n}=\frac{\dd^n}{\dd t^n}$, with integer knots, i.e., we have
\begin{equation} \label{eq:sparse_spline}
	\mathrm{D}^{n}\{\beta_{n-1}\}(t) = \sum_{k\in\bb Z} d^{n}_k \delta(t - k),
\end{equation}
where $\delta(\cdot)$ denotes the Dirac impulse, and $\bld d^{n} = \{d_k^n\}_{k\in\bb Z}$ is the $n^\text{th}$-order finite-difference sequence characterized by its $\cl Z$-transform $(1-z^{-1})^n$ and has $(n+1)$ nonzero entries. For instance, $\bld d^1 = \{\boxed{1}, -1\}$, where $\boxed{\cdot}$ denotes the zero-indexed entry, $\bld d^2 = \{\boxed{1}, -2, 1\}$ is obtained by the linear convolution $\bld d^1 *\bld d^1$, and in general, we have $\bld d^n = \bld d^{n-1}*\bld d^1$. Explicitly, $d_k^n = (-1)^{k}{n\choose k}, k\in\llbracket 0,n+1\rrbracket$.

\subsection{Integer Shift-Invariant Spaces}
\label{subsec:int_siss}
Let $\varphi\in L_2(\bb R)$ denote a {\it generator kernel} that gives rise to the {\it shift-invariant space} defined as
\begin{equation}\label{eq:integer_siss}
	\cl V(\varphi) = \left\{ f: f(t) = \sum_{k\in\bb Z} c_k \varphi \left(t-k\right), \bld c \in\ell_2(\bb Z)\right\}.
\end{equation}
The sequence $\bld c = \{c_k\}_{k\in\bb Z}\in \ell_2(\bb Z)$ denotes the {\it expansion coefficients} of $f$. Additionally, we consider generator kernels to come from $\cl C^0(\bb R)$. Define $\mathrm{V}_\varphi : \ell_2(\bb Z) \rightarrow \mathcal V(\varphi)$ as the synthesis operator that maps the expansion coefficients to a continuous-time signal. The bases $\{\varphi(\cdot-k)\}_{k \in \bb Z}$ are said to constitute a Riesz bases of $\cl V(\varphi)$ if they are linearly independent and the following partial energy-equivalence condition is satisfied:
\begin{equation} \label{eq:riesz_bounds}
	A_{\varphi} \norm{\bld c}_{\ell_2} \leq \norm{f}_{L_2} \leq B_{\varphi}\norm{\bld c}_{\ell_2},
\end{equation}
where $A_\varphi$ and $B_\varphi$ are the lower and upper Riesz bounds, respectively, given in terms of the square-root of the spectrum of the sampled autocorrelation of the generator kernel $\varphi$, denoted by $G_{\varphi}(\omega)$ and defined as follows:
\begin{equation}\label{eq:sampled_autocorrelation_spectrum}
\begin{split}
G_{\varphi}(\omega) \triangleq \sqrt{\sum_{k\in\bb Z}\left|\hat{\varphi}\left(\omega + 2\pi k\right) \right|^2},&\\
A_\varphi = \inf_{\omega \in \left[-\pi,\pi\right]} G_{\varphi}(\omega), \;
B_\varphi = \sup_{\omega \in \left[-\pi,\pi\right]} & G_{\varphi}(\omega).
\end{split}
\end{equation}


\subsection{$h$-Shift-Invariant Spaces}
\label{subsec:h_siss}
\indent In deriving sharp estimates on the approximation quality of a certain Riesz bases, it is useful to consider scaled versions of the generator kernel, i.e., $\varphi_h(t) = \varphi(t/h)$, $h>0$, and the spaces generated by $\varphi(\cdot-hk)$, which gives a direct control over the resolution of the approximation.\\ 
\indent The prototypical $h$-shift-invariant space is the Paley-Wiener space of finite-energy bandlimited functions given by 
\begin{equation*}\label{eq:bandlimited_model}
\begin{split}
	\cl{PW}_{[-\frac{\pi}{h},\frac{\pi}{h}]}\!=\!\cl V(\text{sinc}_h)\!&=\!\left\{ f\!\in\! L_2(\bb R) \!: \! \text{supp}(\hat{f}) \subseteq \left[ -\frac{\pi}{h}, \frac{\pi}{h}\right] \right\},
\end{split}
\end{equation*}
where $\sinc_h(t) = \sinc(t/h)$ denotes the $h$-scaled generator kernel. The set $\left\{\text{sinc}_h\left(\cdot-hk\right)\right\}_{k\in\bb Z}$ constitutes on orthogonal basis for $\cl V(\text{sinc}_h)$ \cite{christensen2010functions}, and therefore, the Riesz bounds are
\begin{equation} \label{eq:sinc_orthogonality}
    A_{\sinc_h} = B_{\sinc_h} = \sqrt{h}.
\end{equation}
\indent The $h$-scaled B-spline of degree $n$, for scale-factor $h>0$ is denoted as $\beta_{n,h}(t) =\beta_n\left(t/h\right)$. The $h$-scaled polynomial B-splines have compact support, more precisely, $\text{supp}(\beta_{n,h}) = [0,h(n+1)]$, and the size of the support depends only on the degree $n$ and scale $h$. The $h$-scaled B-spline of degree $n$ and its translates $\mathcal{B}_{n,h} = \left\{\beta_{n,h}\left(\cdot-hk\right)\right\}_{k\in\bb Z}$ generate the $h$-shift-invariant space
\begin{equation*}\label{eq:spline_model}
	\mathcal V(\beta_{n,h})\! =\! \left\{ f : f(t) \!=\! \sum_{k\in\bb Z} c_k \beta_{n,h}(t-hk), \bld c\in\ell_2(\bb Z)\right\}.
\end{equation*}
For $\varphi = \beta_{n,h}$, the Riesz bounds satisfy \cite[Chapter~10]{christensen2010functions}
\begin{align}\label{eq:spline_riesz_basis}
    \left(\frac{2h}{\pi}\right)^{n+1} \leq A_{\beta_{n,h}},\; \text{ and } B_{\beta_{n,h}} = h^{n+1}.
\end{align}
\indent The orthogonal projection operator onto $\cl V(\varphi)$, denoted as $\Pi_\varphi : L_2(\bb R)\rightarrow \cl V(\varphi)$, is given by
\begin{equation}\label{eq:projection_operator}
	\Pi_{\varphi}\{f\}(t) = \sum_{k\in\bb Z} \langle f, \tilde{\varphi}(\cdot-k)\rangle \;\varphi(t-k),
\end{equation}
where $\tilde{\varphi}\in\cl V(\varphi)$ is the biorthogonal counterpart of $\varphi$ that satisfies
\[
	\langle \tilde{\varphi}, \varphi(\cdot-k)\rangle = \begin{cases}
		1, & k=0\\
		0, & \text{otherwise}
	\end{cases}.
\]
The biorthogonal kernel $\tilde{\varphi}$ has a simpler expression in the Fourier domain: $\displaystyle \hat{\tilde{\varphi}}(\omega) = \frac{\hat{\varphi}(\omega)}{G_\varphi(\omega)}$, where $\hat{\tilde{\varphi}}$ denotes the Fourier transform of $\tilde{\varphi}$.\\
\indent From a practical standpoint, signals having compact support are of greater relevance and interest. Suppose $\varphi$ has compact support $[0,T]$, then
\begin{equation*}\label{eq:compact_spline_model}
	\mathcal V_K(\varphi)\! =\! \left\{ f : f(t) \!=\! \sum_{k=0}^{K-1} c_k \varphi(t-k)\right\},
\end{equation*}
is the counterpart of $\cl V(\varphi)$ containing signals compactly supported over $[0, K-1+T]$. Signals in $\cl V_K(\beta_{n,h})\subset L_2([0,(K+n)h])$ have support $[0,(K+n)h]$, and are completely described by $K$ expansion coefficients $\{c_k\}_{k=0}^{K-1}$. The compactly supported signal model is preferred in image and video processing applications \cite{unser1999splines,arigovindan2005variational}.


\subsection{Time-Encoding Machines}
\label{subsec:time_encoding_prelims}
A time-encoding machine (TEM) may be viewed as an operator that maps continuous-time signals to a sequence of temporal measurements. Gontier and Vetterli provided formal definitions of IF-TEM and C-TEM \cite{gontier2014sampling}, which we unify to provide the following formal definition of a TEM.
\begin{definition}[Time-encoding machine]\label{def:tem}
	Let $\sE$ be an operator on $L_2(\bb R)$ and let $\{r_{i}\}_{i\in\bb N}$ be a set of \emph{reference} functions. The associated time-encoding machine is a map $\sT : f \mapsto \sT\{f\}$, with the following properties:
	\begin{enumerate}[label=\alph*.]
		\item $\sT\{f\} = \{t_i  : t_i > t_j, \; \forall i>j, i \in \bb N\}$,
		\item $\displaystyle \lim_{m\rightarrow\pm\infty} t_{m} = \pm \infty$,\, \text{and}
		\item $\sE\{f\}(t_m) = r_{m}(t_{m}), \forall t_{m} \in \sT\{f\}$.
	\end{enumerate}
\end{definition}
In Section~\ref{sec:event_generation}, we show that the neuromorphic sensor can be modelled as a TEM with the identity event operator (cf. Definition~\ref{def:neuromorphic_encoder}).\\
\indent In the context of time-encoding, we are interested in the so-called $t$-transform \cite{lazar2003recovery} that maps the temporal measurements $\{t_m\}$ to the amplitude samples $\{\sE\{f\}(t_m)\}$. The $t$-transform enables one to solve the signal reconstruction problem in time-encoding within the framework of nonuniform sampling \cite{marvasti2012nonuniform} A formal definition of the $t$-transform follows.
\begin{definition}[$t$-transform]
The $t$-transform of a time-encoding machine is a mapping
\[
	\sT\{f\} \mapsto \{\sE\{f\}(t_m)\}_{m\in\bb N}.
\]
\end{definition}
The $t$-transform can be derived from the event generation process of the time-encoding mechanism. We will derive the $t$-transform for neuromorphic sampling in Lemma~\ref{lem:ttransform}.


\section{Neuromorphic Encoding Model}
\label{sec:event_generation}
\begin{figure*}[!t]
	\centering
	\includegraphics[width=6.5in]{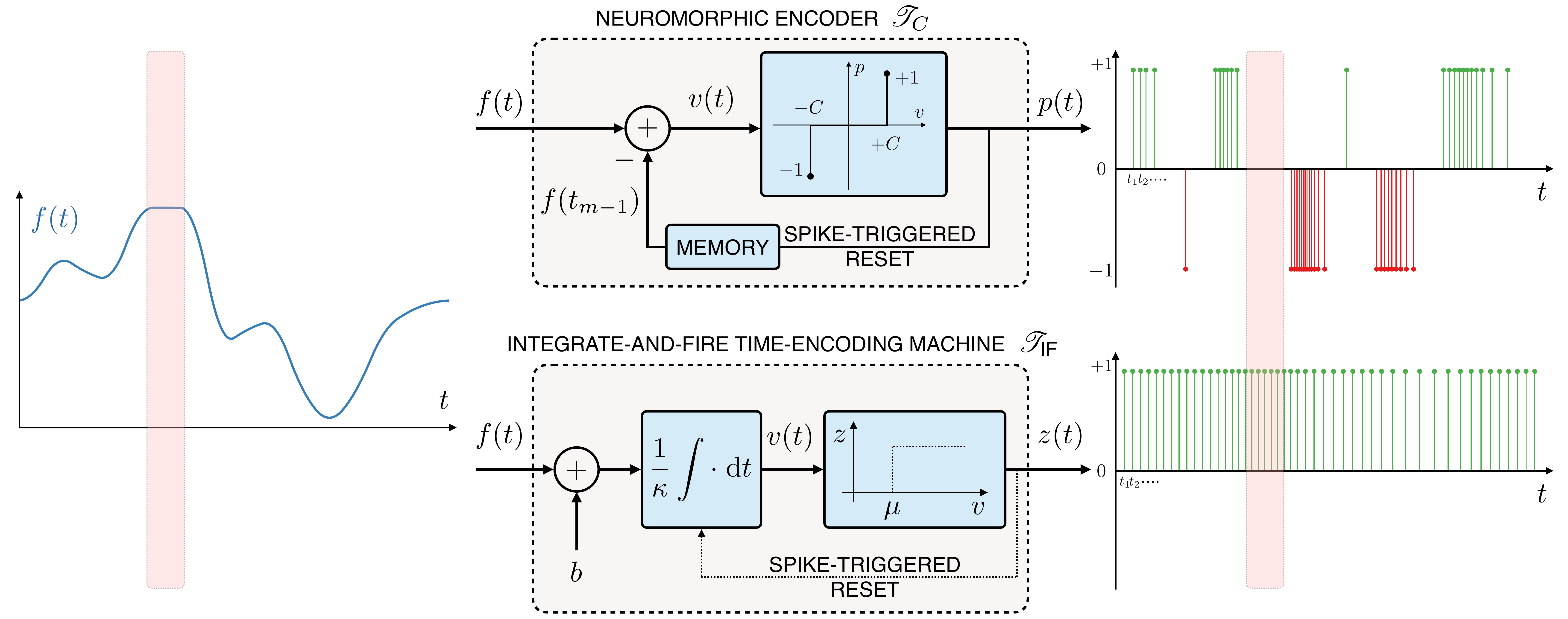}
	\caption{[Top] Schematic of a neuromorphic encoder with temporal contrast threshold $C$. The input $f(t)$ is offset by $f(t_{m-1})$ and compared against $\pm C$. When the difference exceeds (subceeds) $+C$ ($-C$), a spike is given at the output, and the spike-triggered reset control resets the memory to the new amplitude value. The output signal $p(t)$ is a spike-train at locations $t_m$ with polarity $p_m\in\{-1,+1\}$, $m\in\bb Z$. An event is the 2-tuple $(t_m,p_m)$. [Bottom] Schematic of an IF-TEM, and its output signal $z(t)$. The interval highlighted (red) indicates the interval of inactivity in $f(t)$. Although both neuromorphic and IF-TEM measurements are sparse, the former is relatively sparser.}
	\label{fig:schematic_sampler}
\end{figure*}
We present a model for neuromorphic sampling of continuous-time signals. We make a connection between neuromorphic sampling and time-encoding. A two-dimensional array of neuromorphic sensors behaves like an event camera \cite{lichtsteiner2008sensor,brandii2014sensor,gallego2022event} and generates spatially independent event streams.\\
\indent A neuromorphic encoder maps a continuous-time signal to a sequence of $2$-tuples of the type $(t_m, p_m)\in \bb R\times \{-1,+1\}$, referred to as an {\it event}, where $t_m$ is the {\it event time instant} at which the signal changes in magnitude by the temporal contrast threshold $C$, and $p_m$ is the {\it event polarity} that indicates the sign of the change. Based on this operating principle, we introduce a formal definition for the neuromorphic encoder.
\begin{definition}[Neuromorphic encoder]\label{def:neuromorphic_encoder}
Let $f(t)$ be the input to a neuromorphic encoder with temporal contrast threshold $C>0$. The output of the encoder is the set of measurements, $f \mapsto \sT_C\{f\} = \{(t_m,p_m)\}_{m\in\bb N}$, such that
\begin{equation}
\begin{split}
	t_{m} &= \min_{t>t_{m-1}} \left\{t : |f(t) - f(t_{m-1})| = C\right\}, \\
	p_m &= \text{sgn}(f(t_m) - f(t_{m-1})),
\end{split}
\end{equation}
where $\text{sgn}(\cdot)$ denotes the signum function.
\end{definition}
Figure~\ref{fig:schematic_sampler} shows a schematic of the neuromorphic encoder implemented using a comparator. The comparator accepts the difference between the input continuous-time signal $f(t)$ and the amplitude sample $f(t_{m-1})$ evaluated at the previous event time-instant $t_{m-1}$ against thresholds $\pm C$. If the difference exceeds $+C$ at time instant $t=t_m$, a positive spike is produced at the output (an ON event). Likewise, if the difference subceeds $-C$ at time instant $t=t_m$, a negative spike is produced at the output (an OFF event). The output of the comparator is a continuous-time signal of the form
\[
	p(t) = \sum_{m\in\bb N} p_m\delta(t-t_m),
\]
which is a spike train with binary amplitudes $\{p_m\}$ indicating the polarities at locations $\{t_m\}$. We assume that the output of the encoder is the set of $2$-tuples $\{(t_m,p_m)\}_{m\in\bb N}$, although, in principle, additional processing is required to determine the polarities and time instants (for instance, by using sub-Nyquist methods \cite{seelamantula2010sub,vetterli2002sampling} or by deploying a counter \cite{brandii2014sensor,naaman2021temhardware}).\\
\indent The {\it event time-instants} $\{t_m\}_{m\in\bb N}$ are typically nonuniformly spaced. The counterpart of sampling interval in the context of nonuniform sampling is the so-called {\it sampling density} or {\it Beurling density} \cite{landau1967necessary}, which is defined next.
\begin{definition}[Sampling density]
The sampling density of a sequence $\{t_m\}_{m\in\bb N}$ with nondecreasing entries is defined as
\begin{equation}\label{eq:sampling_density}
	\mathfrak{D}(\{t_m\}_{m\in\bb N}) \triangleq \sup_{m\in\bb N} |t_{m+1}-t_{m}|.
\end{equation}
The local sampling density of a finite sequence $\{t_m\}_{m=0}^M$ with nondecreasing entries is defined as
\begin{equation}\label{eq:local_sampling_density}
	\mathfrak{D}(\{t_m\}_{m=0}^{M}) \triangleq \sup_{m=1,\ldots,M} |t_{m+1}-t_{m}|.
\end{equation}
\end{definition}
\indent The neuromorphic sampling mechanism ensures that there are no events in intervals of inactivity in the input signal, as opposed to an IF-TEM \cite{lazar2004period}. Figure~\ref{fig:schematic_sampler} compares a neuromorphic encoder with an IF-TEM. The portion shaded in red indicates the interval of inactivity in the signal, in which, the IF-TEM continues to produce measurements. This is due to the addition of the bias $b$ to the signal before accumulation. The bias can be controlled suitably to produce a set of trigger times with sufficient sampling density. On the other hand, the neuromorphic encoder does not produce any events in the intervals of inactivity. When the signal is a constant, no events will be generated for any choice of the temporal contrast threshold. The neuromorphic encoder is essentially an opportunistic sampler that makes compressed acquisitions right at the source \cite{guan2007opportunistic}. A neuromorphic encoder can also be viewed as a generalization of the crossing time-encoding machine \cite{gontier2014sampling}, where an oracle determines the $m^{\text{th}}$ reference signal in an event-index dependent fashion: $f(t_{m-1}) + p_m C$. The additional encoding of the polarities is central to a neuromorphic encoder. A neuromorphic encoder can be viewed as a time-encoding machine with an associated $t$-transform as explained in the following lemma.
\begin{lemma}[Neuromorphic $t$-transform] \label{lem:ttransform}
Consider $f \in \cl C^0(\bb R)$. Let $\sT_C\{f\} = \{(t_m, p_m)\}_{m\in\bb N}$ denote the sequence of events generated by the neuromorphic encoder $\sT_C$ with temporal contrast threshold $C>0$, and an initial value $(t_0, f(t_0))$ be known. The samples of the signal at event time-instants $\{t_m\}_{m\in\bb N}$ are given by
\begin{equation}\label{eq:lemma2}
	f(t_m) = f(t_0) + C\sum_{i=1}^m p_i, \;\forall m\in\bb N.
\end{equation}
\begin{proof}
From Definition \ref{def:neuromorphic_encoder}, the output $\sT_C\{f\}$ satisfies
\begin{equation*}
\begin{split}
C &= |f(t_m) - f(t_{m-1})|, \;\\
p_m &=  \text{sgn}(f(t_m) - f(t_{m-1})).
\end{split}
\end{equation*}
Succinctly put, we have a recursive relation for the samples of the signal: $f(t_m) = f(t_{m-1}) + p_m C$. Solving the recursion with the initial value $f(t_0)$ at $t_0$ gives Eq.~\eqref{eq:lemma2}.
\end{proof}
\end{lemma}
The temporal contrast threshold $C$ is a user-specified parameter and is held fixed throughout the acquisition process. The initial value $(t_0,f(t_0))$ is assumed to be known a priori.\\
\indent The neuromorphic sensing mechanism described above is {\it asynchronous} in the sense that a synchronizing clock is not needed. Extending the idea to a two-dimensional array of pixels, where each pixel follows a neuromorphic sensing mechanism described above, is an accurate model of a dynamic vision sensor. In addition to the polarities and event instants, the ordinate and abscissa of the pixel must also be recorded. This results in the so-called {\it address-event representation} (AER), where a video with spatial dimension $P\times Q$ is encoded in the form of 4-tuples coming from $\llbracket 1,P\rrbracket\times \llbracket 1,Q\rrbracket\times \bb R\times \{-1,+1\}$.

\section{Perfect Reconstruction of Signals in Shift-Invariant Spaces}
\label{sec:perfect_reconstruction}
\begin{figure}[!t]
	\centering
	\includegraphics[width=3.3in]{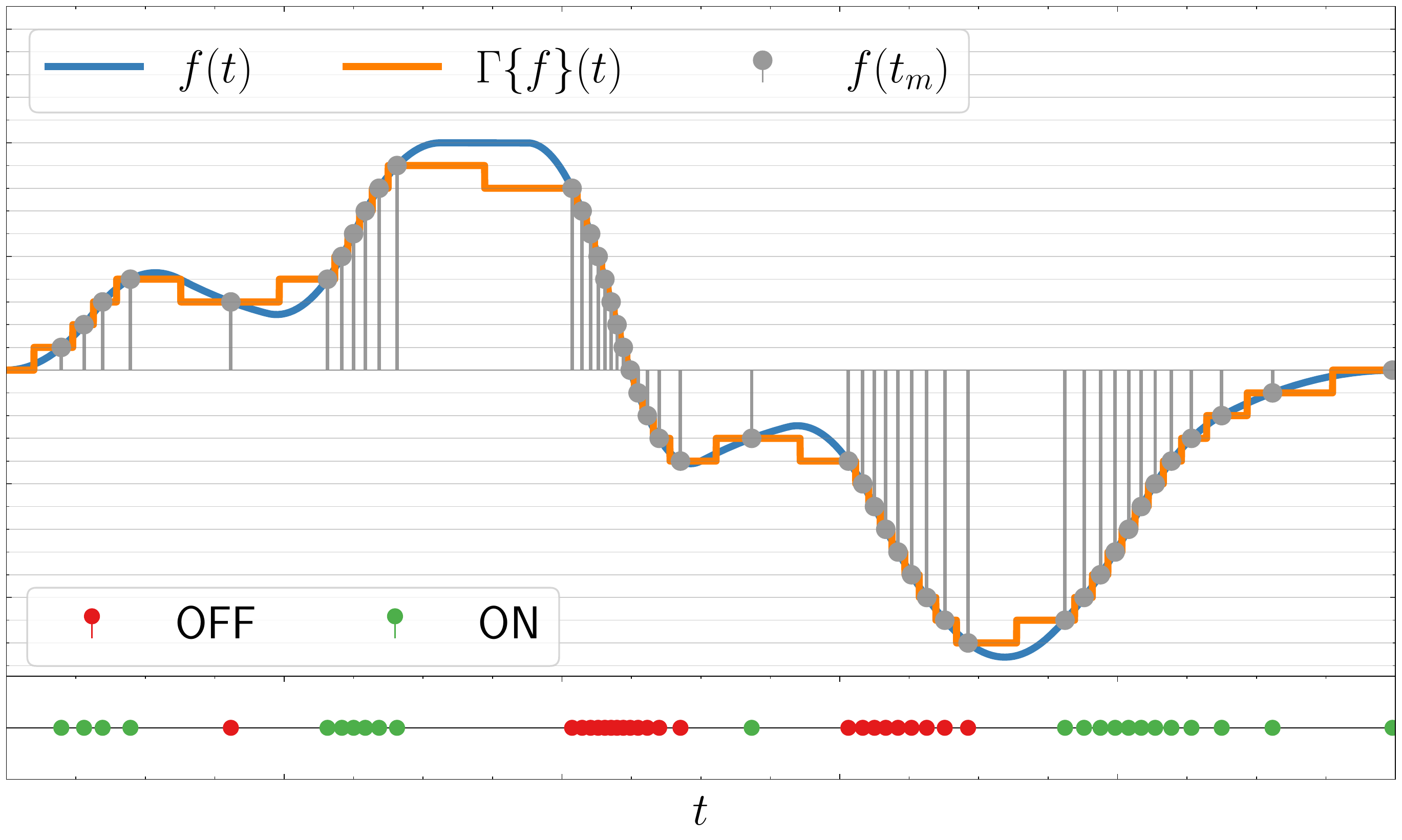}
	\caption{Illustration of neuromorphic sampling and construction of a piecewise-constant approximation $\Gamma\{f\}(t)$ using the samples of $f$ obtained using the $t$-transform. $\Gamma\{f\}(t)$ may not belong to $\cl V(\varphi_h)$, which motivates the need for the alternating projections algorithm for reconstruction.}
	\label{fig:piecewise_constant_approximation_demo}
\end{figure}

In this section, we consider reconstruction of signals in shift-invariant spaces from their events generated using a neuromorphic encoder. The neuromorphic $t$-transform allows one to compute the amplitude samples of the signal from its events. The signal recovery problem is similar to that of reconstruction from nonuniform samples \cite{gontier2014sampling,aldroubi2001nonuniform}. The strategy is to alternate between constructing a piecewise-constant approximation using the amplitude samples, and projecting the approximation to a shift-invariant space containing the signal, until convergence is achieved, i.e., both conditions are satisfied.\\
\indent A piecewise-constant approximation of a function $f$ is constructed from its amplitude samples obtained at event instants $\{t_m\}_{m\in\bb N}$ as follows:
\begin{equation}\label{eq:piecewise_constant_interpolator}
	\Gamma\{f\}(t) = \sum_{m\in\bb N} f(t_m) \mathbbm{1}_{[s_m,s_{m+1}]}(t),
\end{equation}
where $\displaystyle s_m = \frac{t_{m-1}+t_m}{2}$ is the average between consecutive instants, and
\[
	\mathbbm{1}_{[a,b]}(t) = \begin{cases}
		1, & t\in[a,b],\\
		0, & \text{otherwise},
	\end{cases}
\]
denotes the indicator function of the set $[a,b]$. Since the amplitude samples of $f$ are tractable using Lemma~\ref{lem:ttransform}, we can define an approximation to $f$ by projecting the piecewise-constant approximation onto $\cl V(\varphi)$ as $f_{1} = (\Pi_\varphi \Gamma)\{f\}$. The signal $\Gamma\{f\}(t)$ is the nearest-neighbour, piecewise-constant approximation of $f$, given its nonuniform amplitude samples. Figure~\ref{fig:piecewise_constant_approximation_demo} shows an illustration of a continuous-time signal $f(t)$, the ON and OFF events obtained using a neuromorphic encoder, the amplitude samples obtained using the $t$-transform, and the piecewise-constant approximation obtained using Eq.~\eqref{eq:piecewise_constant_interpolator}.


\subsection{Reconstruction as a Continuous-Domain Inverse Problem}
The problem of reconstructing $f$ from its piecewise-constant approximation $f_{1}$ is a continuous-domain linear inverse problem as stated below:
\leqnomode
\begin{equation} \tag{$\bd P_0$}\label{eq:sandberg_problem}
\begin{split}
	\text{Find } f\in\cl V(\varphi_h), \; \text{such that } f_{1} = (\Pi_{\varphi_h} \Gamma)\{f\}.
\end{split}
\end{equation}
\reqnomode
It can be shown, using Sandberg's theorem \cite{sandberg1963properties}, that \eqref{eq:sandberg_problem} has a unique solution given as the limit of the iterations
\begin{equation}\label{eq:sandberg_iterations}
	f_{k+1} = f_{1} + (I - \Pi_{\varphi_h}\Gamma)\{f_{k}\}, \; k=1,2,\ldots.
\end{equation}
The {\it initialization} $f_{1}$ can be computed by projecting $\Gamma\{f\}$ onto $\cl V(\varphi)$, which is tractable since the amplitude samples can be computed using Lemma~\ref{lem:ttransform}. The convergence of the iterates to the true signal depends on the sampling density of the event time-instants $\{t_m\}_{m\in\bb N}$. We recall a result from \cite{gontier2014sampling}, adapted to accommodate signals belonging to $h$-shift-invariant spaces.
\begin{theorem}\label{thm:general_theorem}
Let $\varphi_h, \mathrm{D}\{\varphi_h\} \in L_2(\bb R)$, and $\{t_m\}_{m\in\bb N}$ be a sequence with increasing entries.
Then, $f\in\cl V(\varphi_h)$ can be recovered from $f_{1} = (\Pi_{\varphi_h}\Gamma)\{f\}$ as the limit of $\{f_{k}\}_{k\in\bb N}$ given in Eq.~\ref{eq:sandberg_iterations}, i.e.,
\begin{equation}\label{eq:bandlimited_convergence}
	\norm{f-f_{k}}_{L_2} \leq \gamma^k \norm{f}_{L_2} \overset{k\rightarrow \infty}{\longrightarrow} 0,
\end{equation}
where $\displaystyle\gamma \triangleq \left(\sup_{\omega\in[-\frac{\pi}{h},+\frac{\pi}{h}]}\frac{G_{\mathrm{D}\{\varphi_h\}}(\omega)}{G_{\varphi_h}(\omega)}\right) \frac{\fr D(\{t_m\}_{m\in\bb N})}{\pi} < 1$.
\end{theorem}
The proof is given in Appendix~\ref{appendix:proof_general_theorem}. Convergence can be ensured by obtaining events at sampling density
\[
	\fr D(\{t_m\}_{m\in\bb N}) < \pi\left(\displaystyle\sup_{\omega\in[-\frac{\pi}{h},+\frac{\pi}{h}]}\frac{G_{\mathrm{D}\{\varphi_h\}}(\omega)}{G_{\varphi_h}(\omega)}\right)^{-1} \triangleq \Delta_{\varphi_h}.
\]
It can be shown that, for signals in the Paley-Wiener space $\mathcal{PW}_{[-\frac{\pi}{h},\frac{\pi}{h}]}$, the following property holds:
\[
	\sup_{\omega\in[-\frac{\pi}{h},+\frac{\pi}{h}]}\frac{G_{\mathrm{D}\{\sinc_h\}}(\omega)}{G_{\sinc_h}(\omega)} = \frac{\pi}{h}.	
\]
The iterations in Eq.~\eqref{eq:sandberg_iterations} converge to the true signal when the event time-instants are measured at a sampling density $\fr D(\{t_m\}_{m\in\bb N}) < h$.\\
\indent The necessary condition on $C$ for perfect reconstruction of signals in the Paley-Wiener space $\mathcal{PW}_{[-\frac{\pi}{h},\frac{\pi}{h}]}$ is presented in the following. 
\begin{corollary}[Perfect reconstruction in $\mathcal{PW}_{[-\frac{\pi}{h},\frac{\pi}{h}]}$]\label{cor:bandlimited}
Let $f\in\cl V(\sinc_h)$ and $\sT_C\{f\} = \{(t_m,p_m)\}_{m\in\bb N}$ be the events measured using a neuromorphic encoder with temporal contrast threshold $C>0$. If the iterates given in Eq.~\eqref{eq:sandberg_iterations} converge to $f$, then $C$ must satisfy the following condition: 
\begin{equation}
	C < \frac{\pi}{\sqrt{h}} \norm{f}_{L_2}.
\end{equation}
\end{corollary}
The proof is given in Appendix~\ref{appendix:proof_cor_bandlimited}. Conditions for perfect reconstruction require that every $h$-length interval $[\tau,\tau+h], \forall \tau\in\bb R$ must have at least one measurement.\\
\indent Next, we present a similar condition for perfect reconstruction of signals in $\cl V(\beta_{n,h})$.
\begin{corollary}[Perfect reconstruction in $\cl{V}(\beta_{n,h})$]\label{cor:spline}
Let $f = \mathrm{V}_{\beta_{n,h}}\bld c \in\cl V(\beta_{n,h})$ and $\sT_C\{f\} = \{(t_m,p_m)\}_{m\in\bb N}$ be the events recorded using a neuromorphic encoder with temporal contrast threshold $C>0$. If the iterations given in Eq.~\eqref{eq:sandberg_iterations} converge, then $C < \norm{\bld{c}}_{\ell_1}\eta$,
where $\eta>0$ is a constant that depends only on the degree $n$ of the polynomial B-spline. Further, if $f\in\cl V_K(\beta_{n,h})$, then
\begin{equation*}
	C < \sqrt{K}\left(\frac{\pi}{2h}\right)^{n+1}\eta\norm{f}_{L_2}.
\end{equation*}
\end{corollary}
The proof is provided in Appendix~\ref{appendix:proof_cor_spline}. Conditions for perfect reconstruction require that every interval $[\tau,\tau+h\frac{\eta}{2}], \forall \tau\in\bb R$ of length $\Delta_{\varphi_h}=h\frac{\eta}{2}$ have at least one measurement.


\subsection{Sufficient Conditions for Perfect Reconstruction}
\label{subsec:sufficient_conditions}
The control variable to obtain minimal number of measurements is the temporal contrast threshold $C$. We define the \emph{critical threshold} as
\[
	C_f(\Delta) \triangleq \frac{1}{2} \inf_{\tau\in\bb R} \abs{\max_{\tau<t<\tau+\Delta} f(t) - \min_{\tau<t<\tau+\Delta} f(t)}.
\]
The minimal number of measurements can be obtained by setting the temporal contrast threshold $C$ to a value lower than the critical threshold $C_f(\Delta_\varphi)$. This result is summarized next.
\begin{corollary}[Sufficient condition for perfect reconstruction]\label{cor:sufficient_condition}
	Let $f\in\cl V(\varphi)$ with $C_f(\Delta_{\varphi_h})>0$ and $\sT_C\{f\} = \{(t_m,p_m)\}_{m\in\bb Z}$ be the events measured using a neuromorphic encoder with temporal contrast threshold $C>0$. The signal can be perfectly reconstructed when $C < C_f(\Delta_{\varphi_h})$.
\end{corollary}


\begin{figure*}[t]
	\centering
	\subfigure[Ground truth $f(t)$ and reconstruction $\check{f}(t)$.]{\label{fig:pr_bandlimited_reconstruction}\includegraphics[width=2.75in]{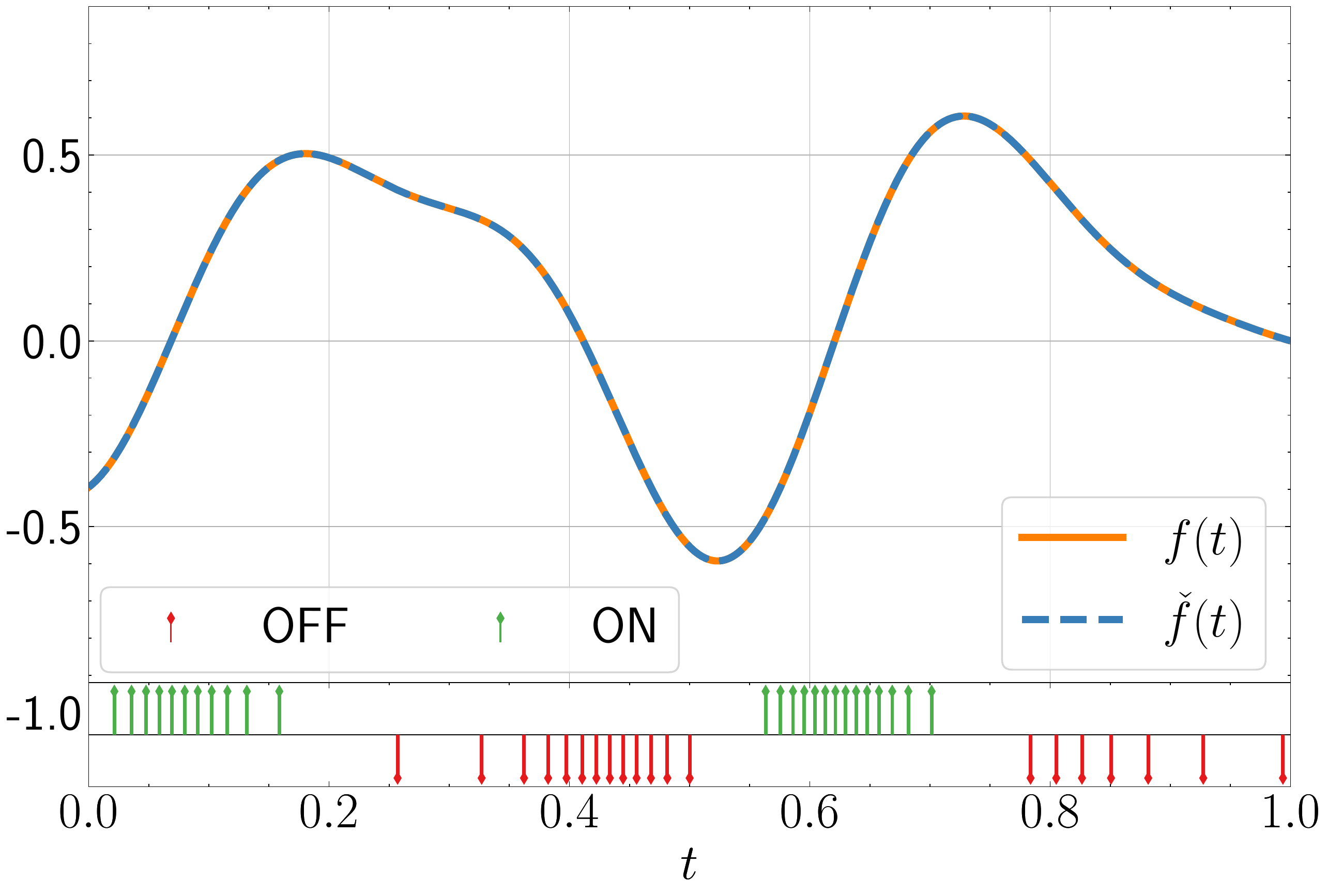}}
	\subfigure[Iterates $\check{f}_{k}(t), k=5,10,15$.]{\label{fig:pr_bandlimited_iterates}\includegraphics[width=2.75in]{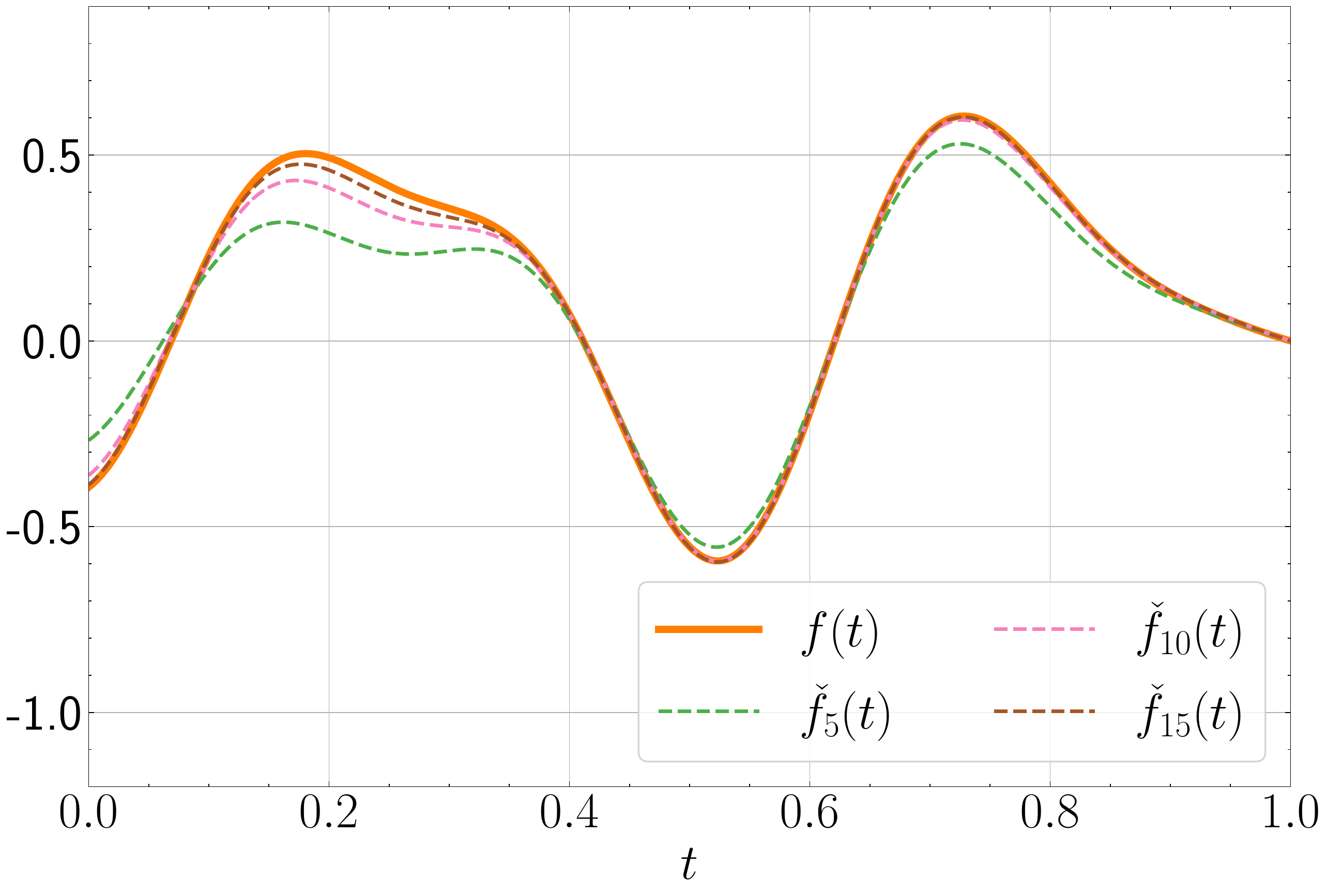}}
	\subfigure[Error signals $e_{k}(t), k=5,10,15$.]{\label{fig:pr_bandlimited_errors}\includegraphics[width=2.75in]{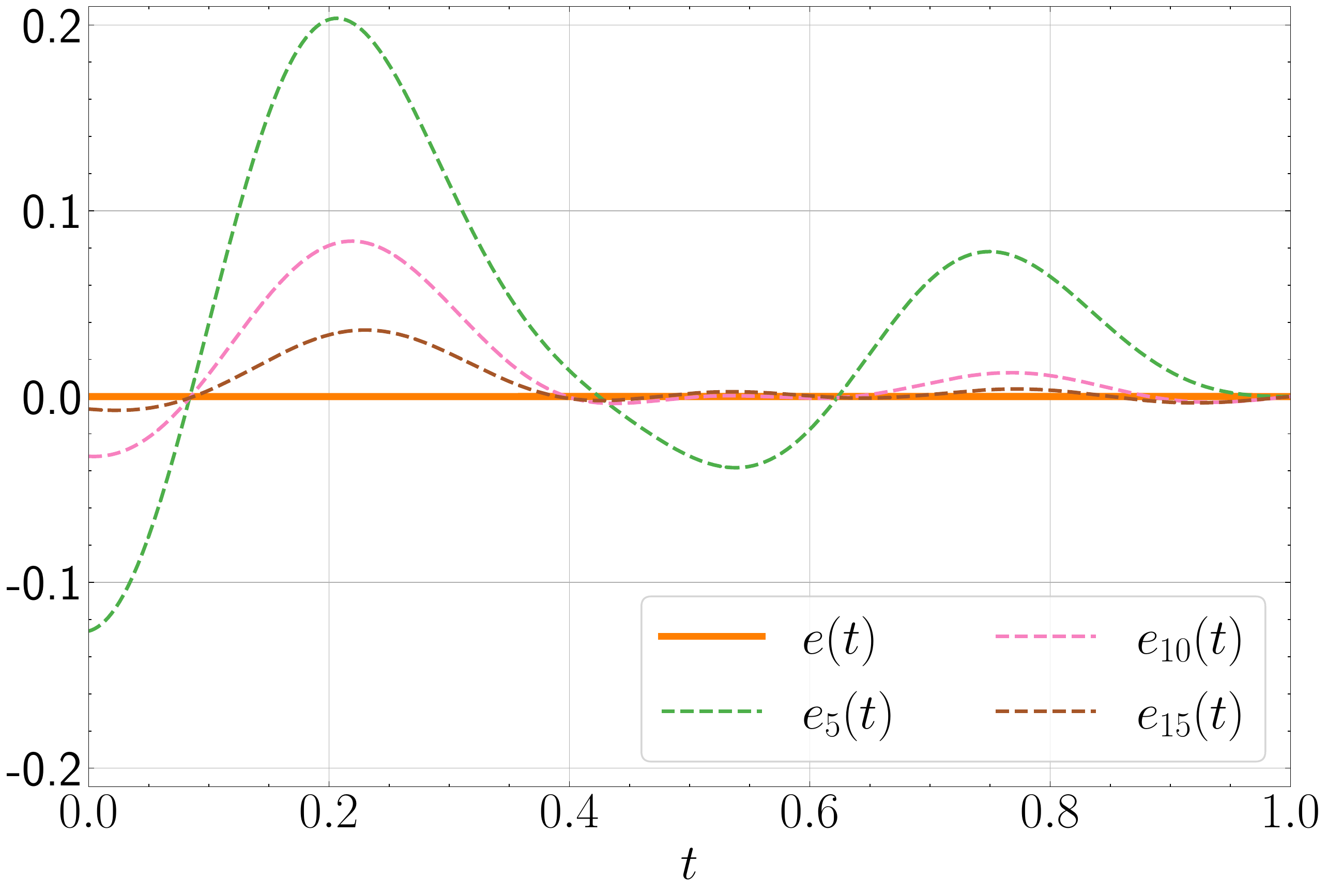}}
	\caption{Neuromorphic sampling and reconstruction of bandlimited signals: (a) ground truth $f(t)$, the events recorded, and the reconstruction $\check{f}(t)$; (b) iterates $\check{f}_{k}(t),\;k=5,10,15$; and (c) error signals $e_{k}(t) = f(t) - \check{f}_{k}(t), \;k=5,10,15$, along with $e(t) = f(t)-\check{f}(t)$. The error signals converge to the zero signal, indicating perfect reconstruction.}
	\label{fig:pr_bandlimited}
\end{figure*}
\begin{figure*}[t]
	\centering
	\subfigure[Ground truth $f(t)$ and reconstruction $\check{f}(t)$.]{\label{fig:pr_splines_converged}\includegraphics[width=2.75in]{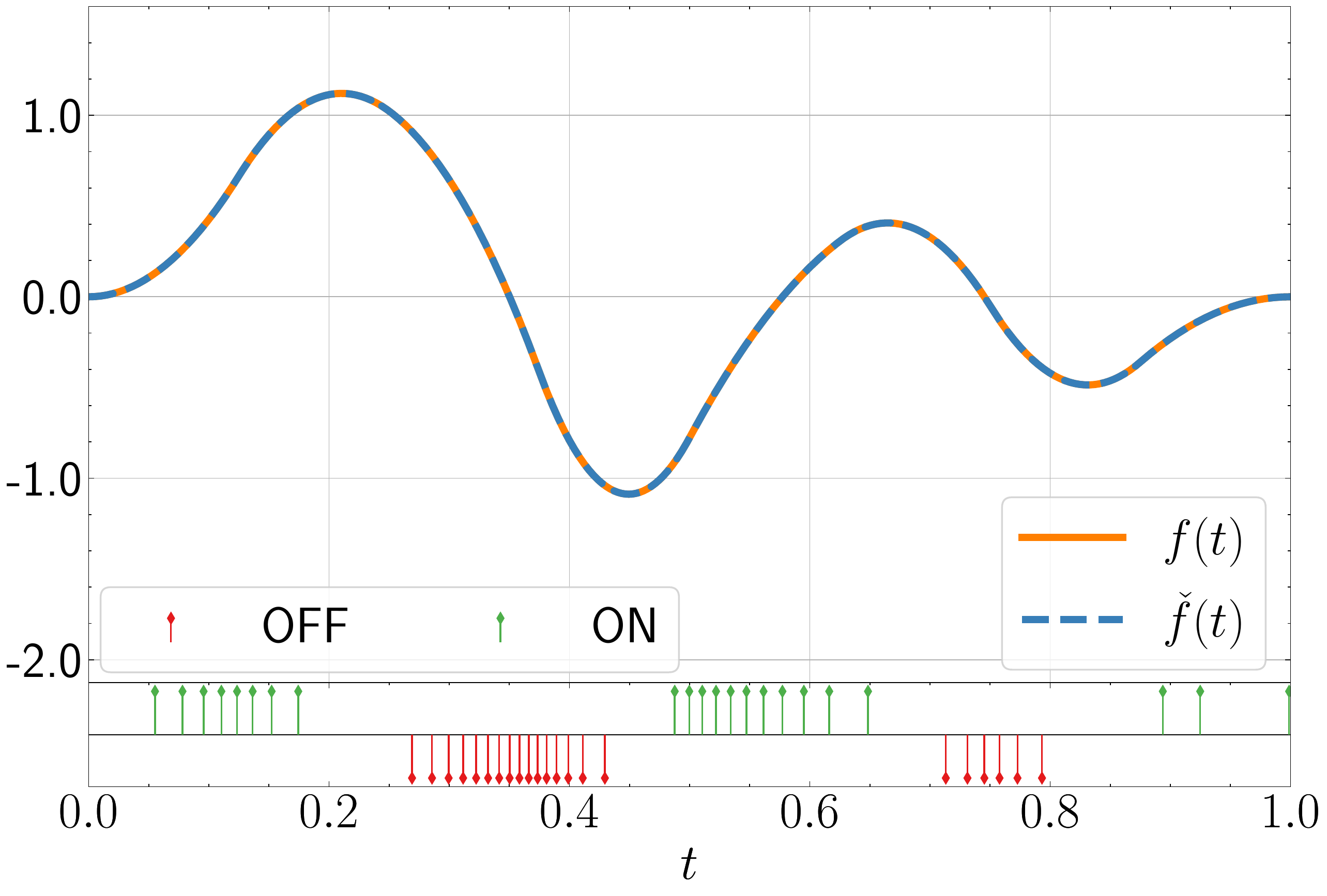}}
	\subfigure[Iterates $\check{f}_{k}(t), k=5,10,15$.]{\label{fig:pr_splines_iterates}\includegraphics[width=2.75in]{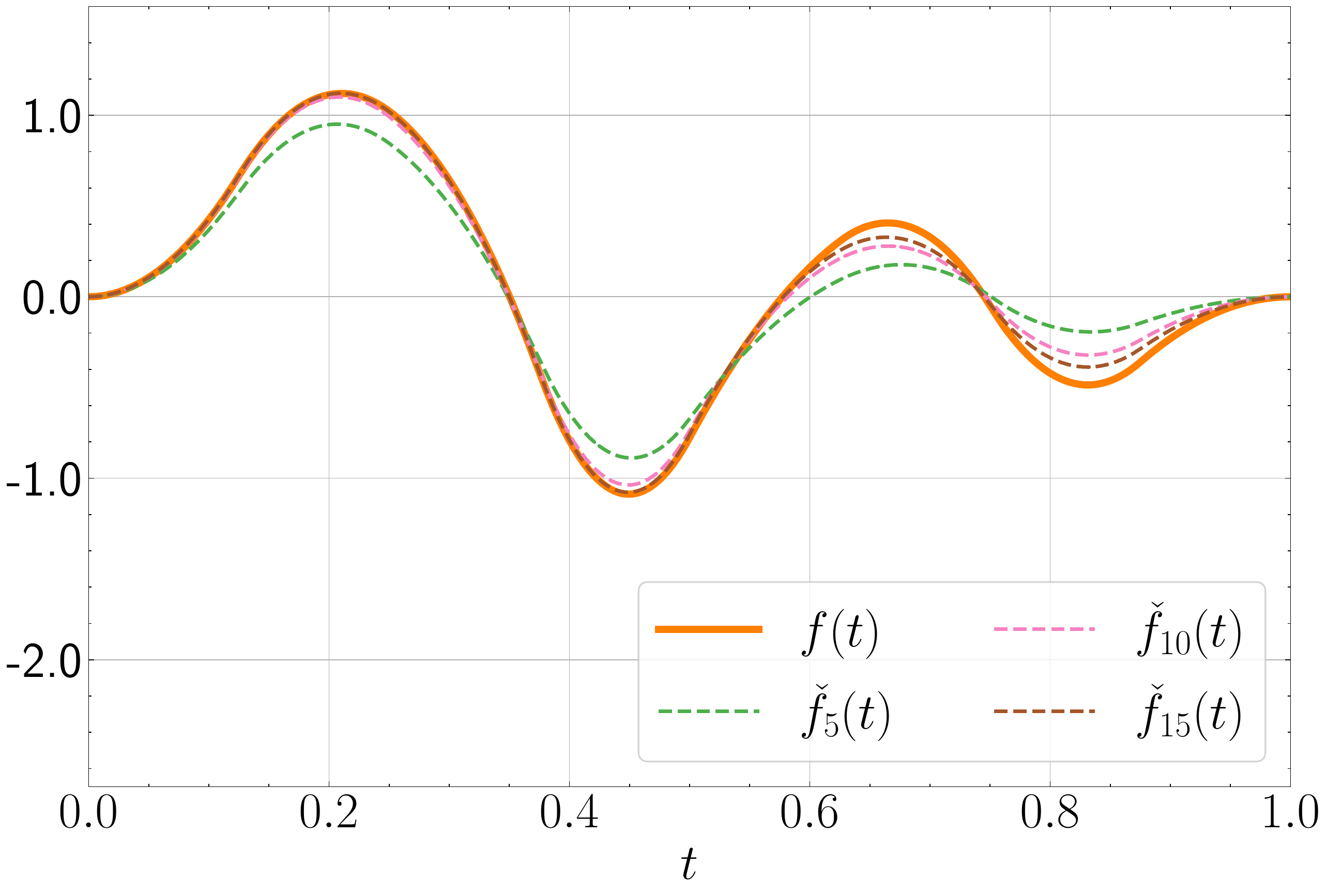}}
	\subfigure[Error signals $e_{k}(t), k=5,10,15$.]{\label{fig:pr_splines_errors}\includegraphics[width=2.75in]{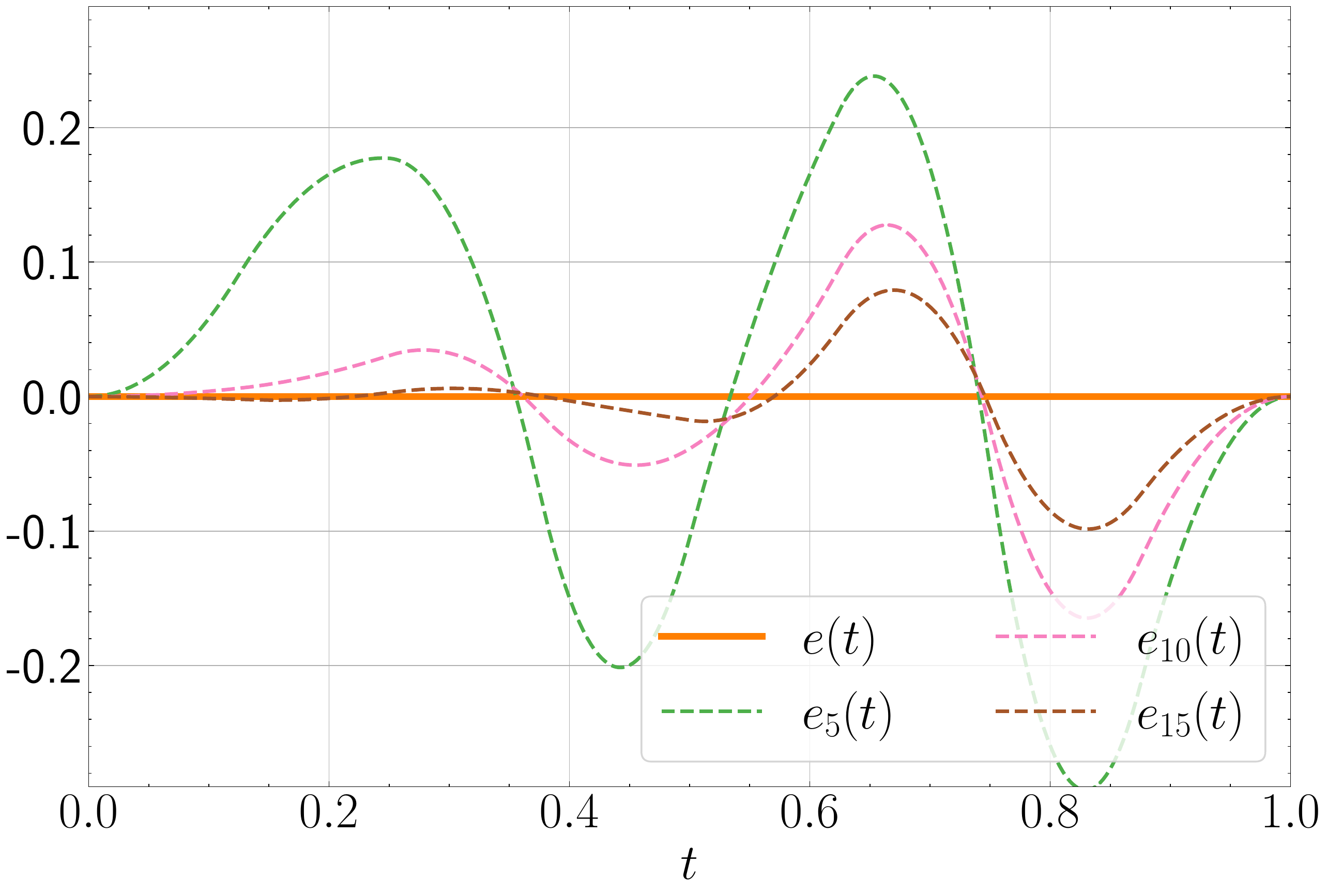}}
	\caption{Neuromorphic sampling and reconstruction of signals in spline spaces of degree $n=2$: (a) ground truth $f(t)$, the events recorded, and the reconstruction $\check{f}(t)$; (b) iterates $\check{f}_{k}(t),\;k=5,10,15$; and (c) error signals $e_{k}(t) = f(t) - \check{f}_{k}(t), \;k=5,10,15$, along with $e(t) = f(t)-\check{f}(t)$. The error signals converge to the zero signal, indicating perfect reconstruction.}
	\label{fig:pr_splines_linear}
\end{figure*}

\subsection{Reconstruction Using Finite-Dimensional Operations}
The continuous-domain iterations in Eq.~\eqref{eq:sandberg_iterations} can be expressed as operations on the expansion coefficients as shown in \cite{gontier2014sampling}. Adapting the result to $h$-shift-invariant spaces, we have the following lemma.
\begin{lemma}\label{lem:finite_updates}
Let $H,\tilde{H} \in \bb R^{\bb Z\times \bb Z}$ defined as $H_{m,k} = \varphi_h (t_m-hk)$ and $\tilde H_{m,k} = \int_{s_m}^{s_{m+1}} \tilde{\varphi}_h (u-hk)~\dd u$, $m\in\bb N,k\in\bb Z$. Suppose $f(t)~=~\sum_{k\in\bb Z} a_k\varphi_h(t-hk)$, then $(\Pi_{\varphi_h}\Gamma)\{f\}(t)~=~\sum_{k\in\bb Z} b_k\varphi_h(t-hk)$, where $\bld b = \tilde{H} H \bld a$.
\end{lemma}
The proof is given in Appendix~\ref{appendix:proof_finite_updates}. Using Lemma~\ref{lem:finite_updates}, the iterations given in Eq.~\eqref{eq:sandberg_iterations} can be expressed in terms of its expansion coefficients as follows:
\begin{equation}\label{eq:finite_iterations}
	\bld c_{k+1} = \bld c_{1} + (I-\tilde{H}H)\bld c_{k},
\end{equation}
where $\bld c_{1} = \tilde{H}\bld y$, with $\bld y = \{f(t_m)\}_{m\in\bb N}$. Condensing the recursion, we write
\begin{equation*}
	\bld c_{k+1} = \sum_{l=0}^{k} (I - \tilde{H}H)^l \bld c_{1},
\end{equation*}
and using Neumann series \cite{horn2012matrix}, it can be seen that the iterates converge to the sequence
\begin{equation}\label{eq:closed_form_equivalent}
	\bld c^* = (\tilde{H}H)^{-1}\tilde{H}\bld y
\end{equation}
\indent From a practical standpoint, we consider reconstruction of compactly supported signals in $\cl V_K(\beta_{n,h})$. We consider signals in $\cl V_K(\varphi)$ that can be expressed by finite expansion coefficients $\bd c = [c_0\; c_1 \cdots c_{K-1}]^\TT\in\bb R^K$. Suppose we have $M$ event time-instants in the interval $[0,T]$, where $T=(K+n)h$. The amplitude samples computed using Lemma~\ref{lem:ttransform}:
\[
	y_m = f(t_0) + C\sum_{i=1}^m p_i.
\]
Denote $\bd y = [y_1\; y_2\cdots y_M]^\TT\in\bb R^M$, which is related to the expansion coefficients $\bd c$ via the matrix $\bd H\in\bb R^{M\times K}$ with entries $\bd H_{m,k} = \varphi_h(t_m-hk)$ as $\bd y = \bd H\bd c$. We recall from \cite{grochenig1993discrete} that the matrix $\bd H$ is left-invertible if the event time-instants are locally dense.
\begin{lemma}[Left-inverse of $\bd H$, \cite{grochenig1993discrete}]\label{lem:leftinverse}
Let $n\in\bb N$, $h>0$ and $\{t_m\}_{m=1}^M$ be a set with increasing entries. Then, the matrix $\bd H\in\bb R^{M\times K}$ with entries $\bd H_{m,k} = \varphi_h(t_m-hk)$ is left-invertible if $\fr D(\{t_m\}_{m=1}^M) < \Delta_{\varphi_h}$.
\end{lemma}

\subsection{Experimental Results}
Consider neuromorphic sampling of a bandlimited signal
\begin{equation*}
	f(t) = \sum_{k=0}^7 f(kh) \;\sinc_h(t-hk),
\end{equation*}
where the uniform samples are chosen as $\{f(kh)\}_{k=0}^7 = \{11.12, 8.58, 7.39, 1.38, 2.08, 2.94, 4.51, 3.89\}$, and $h=2^{-3}$. We obtain neuromorphic measurements with temporal contrast threshold set according to Corollary~\ref{cor:sufficient_condition}. We perform reconstruction using the iterations given in Eq.~\eqref{eq:finite_iterations}, and its closed-form equivalent in Eq.~\ref{eq:closed_form_equivalent}.\\
\indent Figure~\ref{fig:pr_bandlimited_reconstruction} shows the ground-truth signal $f(t)$, the events recorded, and the reconstruction $\check{f}(t)$ obtained using the closed-form solution. Figure~\ref{fig:pr_bandlimited_iterates} shows the signals $\check{f}_{k}$ synthesized using the expansion coefficients iterates $\bld c_{k}, \; k=5,10,15$. We observe that the iterates converge to the ground truth as $k$ increases. Figure~\ref{fig:pr_bandlimited_errors} shows the error signal $e(t) = f(t)-\check{f}(t)$, along with the error signal of the iterates $e_{k}(t) = f(t) - \check{f}_{k}(t)$. We observe that the error signal converges to the zero signal, indicating perfect reconstruction.\\
\indent Next, consider neuromorphic sampling of compactly supported signals over $[0,1]$ in $\cl V_K(\beta_{n,h})$, given as
\begin{equation*}
	f(t) = \sum_{k=0}^{K-1} c_k\beta_{n,h}(t-kh),
\end{equation*}
where $h=2^{-3}$, $n=2$ and the coefficients $\{c_k\}_{k=0}^{K-1}$ are sampled from a standard Gaussian distribution. We obtain neuromorphic measurements with temporal contrast threshold set according to Corollary~\ref{cor:sufficient_condition}, the reconstructions $\check{f}_k(t)$ obtained using Eq.~\eqref{eq:finite_iterations}, and the closed-form solution $\check{f}(t)$ in Eq.~\ref{eq:closed_form_equivalent} (cf. Figure~\ref{fig:pr_splines_linear}). We observe that the iterates converge to the ground truth as $k$ increases. Figure~\ref{fig:pr_splines_errors} shows the error signal $e(t) = f(t)-\check{f}(t)$, and the error in the iterates $e_{k}(t) = f(t) - \check{f}_{k}(t)$. The error signal converges to zero, indicating perfect reconstruction.
The temporal contrast threshold $C$ is set at $\approx 5\%$ of the dynamic range of the signal to ensure that the sampling density conditions are met. This is signal-dependent, and for certain signals, it may not be possible to set a temporal contrast threshold according to Corollary~\ref{cor:sufficient_condition}.


\section{Reconstruction from Overly Sparse Measurements}
\label{sec:sparse_reconstruction}
We now consider reconstruction of signals from overly sparse measurements in the sense that the events do not meet the minimum sampling density criterion of Theorem~\ref{thm:general_theorem}, which gives rise to the possibility of the matrix $\bd H$ becoming rank-deficient. We consider signals in $\cl V_K(\beta_{n,h})$ and extend the result to a larger class of signals by leveraging the universal approximation properties of splines \cite{lei1994lp}. From a practical perspective, we consider the case where the temporal contrast threshold is set to a fraction of the dynamic range of the signal, and may not obey the condition in Corollary~\ref{cor:spline}.\\
\indent Classical approaches in signal reconstruction and interpolation such as Tikhonov regularization \cite{duchon1977splines,arigovindan2005variational,micchelli1984interpolation,meinguet1979multivariate} are popular choices in signal and image processing, and are accompanied by efficient finite-dimensional reconstruction algorithms. We pose signal reconstruction as an $L_p$-(pseudo)norm, $p>0$, minimization problem, similar to \cite{gupta2018continuous,debarre2019b,bohra2020continuous}, and deduce the corresponding discrete-domain optimization problem in terms of the expansion coefficients in $\bb R^K$, which can be efficiently solved using a variable-splitting approach. For certain choices of $p$, for instance, $p\geq 1$, the discrete-domain optimization problem turns out to be convex, for which efficient solvers are available.


\subsection{Reconstruction Using $L_p$ Minimization}
Consider neuromorphic sampling of signals in $\cl V_K(\beta_{n,h})$ of the type $f(t) = \sum_{k=0}^{K-1} c_k \beta_{n,h}(t-hk)$,
which are specified by the expansion coefficients $\bd c = [c_0\;c_1\ldots c_{K-1}]^\TT$. Since the sampling density bounds may not always be met, the reconstruction algorithms proposed in Section~\ref{sec:perfect_reconstruction} cannot be readily deployed. Suppose $f(t)$ is the input to the neuromorphic encoder with temporal contrast threshold $C>0$, and $\sT_C\{f\} = \{(t_m,p_m)\}_{m=1}^{M}$ are $M$ events in the interval $[0,T]$. Let $\bd y = [y_1\; y_2\ldots y_M]^\TT\in\bb R^M$ be the amplitude samples of the signal obtained using Lemma~\ref{lem:ttransform}. The linear system of equations $\bd y = \bd H\bd c$ may be ill-posed. The signal reconstruction can be posed as an interpolation problem. Consider the continuous-domain optimization program
\leqnomode
\begin{equation} \tag{$\bd P_1$}\label{eq:inverse_problem}
\begin{split}
\underset{f\in \cl V_K (\beta_{n,h})}{\text{minimize }} \norm{\mathrm{D}^n\{f\}}_{L_p}, \; \text{subject to } f(t_m) = y_m,
\end{split}
\end{equation}
\reqnomode
where $m\in\llbracket1,M\rrbracket$ and $p>0$. The optimization variable $f$ is constrained to lie in $\cl V_K(\beta_{n,h})$. Bohra and Unser \cite{bohra2020continuous} and Unser \cite{unser2017splines} showed an extension of this formulation with $p\geq 1$ for solving the interpolation problem for a larger class of signals (the space of finite Radon measures). They showed that the solution lies in $\cl V_K(\beta_{n,h})$ for a suitable choice of $h>0$. The objective function with choice of $p =1, 2,$ and $\infty,$ imposes sparsity, smoothness, and boundedness on the derivatives of $f$, respectively. Bohra and Unser \cite{bohra2020continuous} showed that the maximum overshoot in interpolation, i.e., $\norm{f-\check f}_{\infty}$, where $\check f$ is the solution of \eqref{eq:inverse_problem}, decreases with decreasing $p$. This motivates our extension to include $0<p<1$, for which the $L_p$-pseudonorm is a sparsity-promoting objective function.

\subsection{Finite-Dimensional Reconstruction Algorithm}
\label{subsec:lp_algorithms}
\begin{figure}[!t]
	\centering
	\includegraphics[width=2.75in]{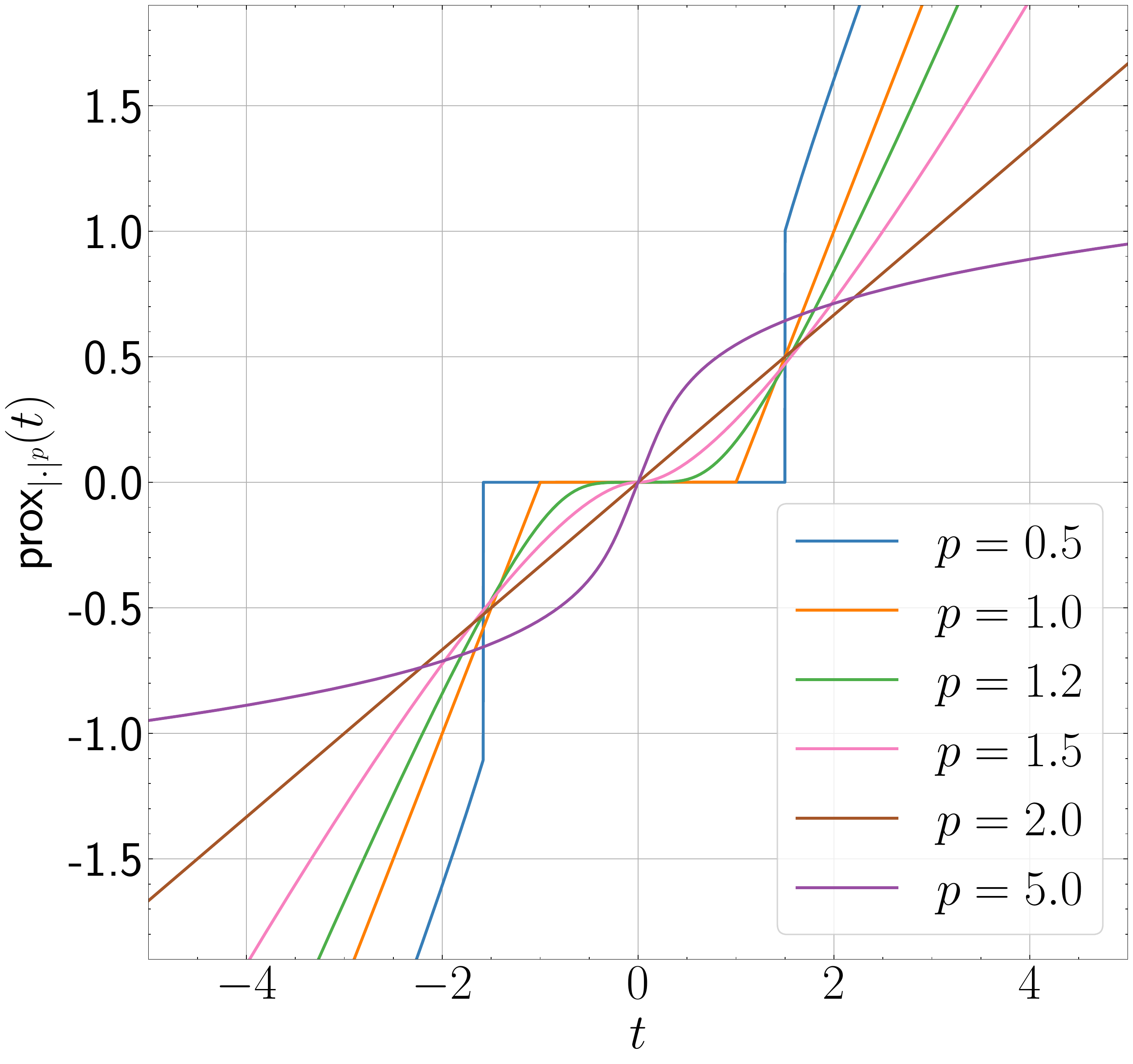}
	\caption{Scalar proximal operator $\text{prox}_{\abs{\cdot}^p}(t)$ for various choices of $p$. For $p=1$, the proximal operator is given by the soft-thresholding operator, and for $p=2$, the proximal operator is linear.}
	\label{fig:prox_ops_demo}
\end{figure}
We develop a finite-dimensional reconstruction algorithm to solve \eqref{eq:inverse_problem}. By restricting the program to signals in $\cl V_K(\beta_{n,h})$, the equality constraint can be readily put in a matrix form as follows: $\bd H\bd c = \bd y$, where $\bd H\in\bb R^{M\times K}$ has entries given by $\bd H_{m,k} = \beta_{n,h}(t_m-hk)$. The objective function is given by
\begin{equation}\label{eq:lp_objective}
	\norm{\mathrm{D}^n\{f\}}_{L_p} = \norm{\frac{1}{h^{n-1/p}}(\bld d^n * \bld c)}_{\ell_p}, \; \forall p>0,n\in\bb N,
\end{equation}
which follows from \cite[Proposition~1]{bohra2020continuous}. Correspondingly, we have the equivalent discrete-domain optimization program:
\leqnomode
\begin{equation} \tag{$\bd P_2$}\label{eq:gTV}
\begin{split}
	\underset{\bd c\in\bb R^K}{\text{minimize }} \Vert \mathbf{Lc}\Vert^p_{p},\quad
	\text{subject to } \mathbf{Hc}=\bd y,
\end{split}
\end{equation}
\reqnomode
where $\bd L\in\bb R^{(K+n)\times K}$ is the discrete linear-convolution matrix corresponding to the sequence $\frac{1}{h^{n-1/p}}\bld d^n$. The program \eqref{eq:gTV} is convex when $p\geq 1$. The program with $p\leq 1$ has a sparsity-promoting objective on $\bd L\bd c$, that effectively enforces sparsity on the $n^\text{th}$-order finite-difference of $\bd c$. The case with $p=1$ is a generalization of analysis basis-pursuit (ABP) or analysis least absolute shrinkage and selection operator (ALASSO), which is a familiar problem in the context of analysis-sparse signal recovery \cite{candes2011compressed,nam2013cosparse,tan2014smoothing}, with the $n^\text{th}$ order finite-difference acting as the analysis operator. In general, since the objective in \eqref{eq:gTV} is not separable in the entries of the optimization variable $\bd c$, we reformulate the optimization with variable-splitting \cite{boyd2004optimisation,boyd2011distributed}
\leqnomode
\begin{equation} \tag{$\bd P_3$} \label{eq:gTV_variable_splitting}
\begin{split}
	\underset{\bd c\in\bb R^K \bd z\in\bb R^{K+n}}{\text{minimize }} \norm{\mathbf{\bd z}}^p_p,\quad
	\text{subject to } \bd L\bd c = \bd z,\;
	\bd H\bd c = \bd y,
\end{split}
\end{equation}
\reqnomode
whose minimizer $(\bd c^*, \bd z^*)$ also gives the minimizer of \eqref{eq:gTV} viz. $\bd c^*$, and hence \eqref{eq:gTV} and \eqref{eq:gTV_variable_splitting} are equivalent. The augmented Lagrangian of \eqref{eq:gTV_variable_splitting} is given by
\begin{equation*} \label{eq:augmented_lagrangian_gtv}
\begin{split}
	 \cl L_p(\bd c, \bd z, \bd u, \bd v)\! =\! \frac{\rho}{2}\Vert \bd y-\mathbf{Hc}\Vert_{2}^2\! +\! \rho\bd v^\TT(\bd H\bd c - \bd y) + & \norm{\bd z}^p_p\\
	+ \lambda\bd u^\TT(\bd L\bd c - \bd z) + \frac{\lambda}{2}\norm{\bd L\bd c - \bd z}^2_{2},&
\end{split}
\end{equation*}
where $\lambda,\rho>0$ are the regularization parameters and $\bd u\in\bb R^{K+n}, \bd v\in\bb R^{M}$ are the dual variables. Using the augmented Lagrangian, we observe that, with  $\rho\rightarrow\infty$, \eqref{eq:gTV_variable_splitting} reduces to \eqref{eq:gTV}. The minimizer of \eqref{eq:gTV_variable_splitting} can be obtained using the updates
\begin{equation*}\label{eq:gtv_iterations}
\begin{split}
	\bd c_{k+1} &= \arg\min_{\bd c\in\bb R^K} \; \cl L_p(\bd c, \bd z_{k}, \bd u_{k}, \bd v_{k}),\\
	&=(\rho\bd H^\TT\bd H\! +\! \lambda \bd L^\TT\bd L)^{-1}\! \left( \rho\bd H^\TT \!(\bd y\!+\!\bd v_{k})\! +\! \lambda \bd L^\TT \!(\bd z_{k}\! + \!\bd u_{k})\!\right),\\
	\bd z_{k+1}\!&=\!\underset{\bd z\in\bb R^{K+n}}{\arg\min}\,\cl L_p(\bd c_{k+1},\!\bd z,\!\bd u_{k},\!\bd v_{k})=\text{prox}_{\frac{\lambda}{\rho}\norm{\cdot}^p_p}\!\left(\bd L\bd c_{k+1}\!+\!\bd u_{k}\right),\\
	\bd u_{k+1}\!&=\!\bd u_{k}\!+\!\nabla_{\bd u}\cl L_p(\bd c_{k+1},\!\bd z_{k+1},\!\bd u,\!\bd v_{k})\!=\!\bd u_{k}\!+\!\bd L\bd c_{k+1}\!-\!\bd z_{k+1},\\
	\bd v_{k+1}\!&=\!\bd v_{k}\!+\!\nabla_{\bd v}\cl L_p(\bd c_{k+1},\!\bd z_{k+1},\!\bd u_{k+1},\!\bd v)\!=\!\bd v_{k}\!+\!\bd H\bd c_{k+1}\!-\!\bd y,
\end{split}
\end{equation*}
where the proximal operator \cite{parikh2014proximal} is given by
\begin{equation*}
\begin{split}
	\text{prox}_{\frac{\lambda}{\rho} \norm{\cdot}^p_p} \left( \bd x\right) &= \arg\min_{\bd z\in\bb R^{K+n}} \frac{1}{2}\norm{\bd x - \bd z}^2_{2} + \frac{\lambda}{\rho}\norm{\bd z}^p_p.
\end{split}
\end{equation*}
\begin{algorithm}[t]
    \caption{Reconstruction of signals in $\cl V_K(\beta_{n,h})$ from events using $L_p$ minimization.} \label{algo:reconstruction_using_lp}
    \KwIn{Measurements $\sT_C\{f\} = \{(t_m, p_m)\}_{m=1}^M$, temporal contrast threshold $C$, grid-size $h>0$, parameters $\lambda, \rho >0$, $p > 0$, degree $n\in\bb N$.}
    {\bf Compute:} $y_m = f(t_0) + C\sum_{i=1}^m p_i, \; m\in \llbracket 1,M\rrbracket$\;
	{\bf Construct:} $\bd H\in\bb R^{M\times K}$ with $\bd H_{m,k} = \beta_{n,h}(t_m-hk)$; $\bd L\in\bb R^{(K+n)\times K}$ as the convolution matrix of $\frac{\bld d^n}{h^{n-1/p}}$ \;
    {\bf Initialize:} $\bd{c} = \bld{0} $, $\bd{z} = \bd{u} = \bld 0, \bd{v} = \bld{0}$ \;
    \SetInd{0em}{.2em}
	\Repeat{\bf convergence or maximum iterations}{
    $\bd c_{k+1}~\!=\!~(\rho\bd H^\TT\bd H\!+\!\lambda\bd L^\TT\bd L)^{\!-\!1}\!\!\left( \rho\bd H^\TT\!(\bd y\!+\!\bd v_{k}\!)\!\!+\!\!\lambda\bd L^\TT\!(\bd z_{k}\!\!+\!\!\bd u_{k}\!)\right)$\;
    $\bd z_{k+1}~=~\text{prox}_{\frac{\lambda}{\rho} \norm{\cdot}^p_p} \left( \bd L\bd c_{k+1} + \bd u_{k}\right)$\;
    $\bd u_{k+1}~=~\bd u_{k} + \bd L \bd c_{k+1} - \bd z_{k+1}$\;
    $\bd v_{k+1}~=~\bd v_{k} + \bd H \bd c_{k+1} - \bd y$
    }
    \KwOut{Reconstruction $\check f = \mathrm{V}_{\beta_{n,h}}\bd c_{k+1}$}
\end{algorithm}
Proximal operators are not always amenable to closed-form evaluation. For $p=1$, the proximal operator is the soft-threshold operator with parameter $\lambda/\rho$. For $p>0$, although a closed-form expression is not available, the proximal operator can be computed by leveraging the component-wise separability property of $\norm{\cdot}_p^p$ as follows:
\begin{align*}
	[\text{prox}_{\frac{\lambda}{\rho}\norm{\cdot}^p_p}\!\left( \bd x\right)]_i\!=\! \text{prox}_{\frac{\lambda}{\rho}\abs{\cdot}^p}\!\left(x_i\right)\!=\!\underset{z\in\bb R}{\arg \min}\frac{1}{2}\left(x_i\!-\!z\right)^2\!+\!\frac{\lambda}{\rho}\abs{z}^p,
\end{align*}
for $i\in\llbracket 1,K\rrbracket$. The scalar optimization problem can be solved relatively easily using standard optimization routines. Figure~\ref{fig:prox_ops_demo} shows the proximal operator for $p>0$. The proximal operators can also be precomputed and stored in the form of a look-up table for ready reference during runtime. Supposing the updates converge to $(\bd c^*, \bd z^*, \bd u^*, \bd v^*)$, the reconstruction can be obtained as $\check f = \mathrm{V}_{\beta_{n,h}}\bd c^*$. The reconstruction technique is summarized in Algorithm~\ref{algo:reconstruction_using_lp}. The algorithm extends to the case where we obtain dense samples by setting $\rho\rightarrow\infty$ and $\lambda=0$, which gives the least-squares solution to $\bd c$ as discussed in Section~\ref{sec:perfect_reconstruction}, although \eqref{eq:gTV} ceases to be a well-posed optimization problem because the feasibility set becomes the null set.
\begin{figure*}[!t]
	\centering
	\subfigure[$n=2$, quadratic polynomial splines]{\label{fig:lp_reconstruction_2}\includegraphics[width=2.75in]{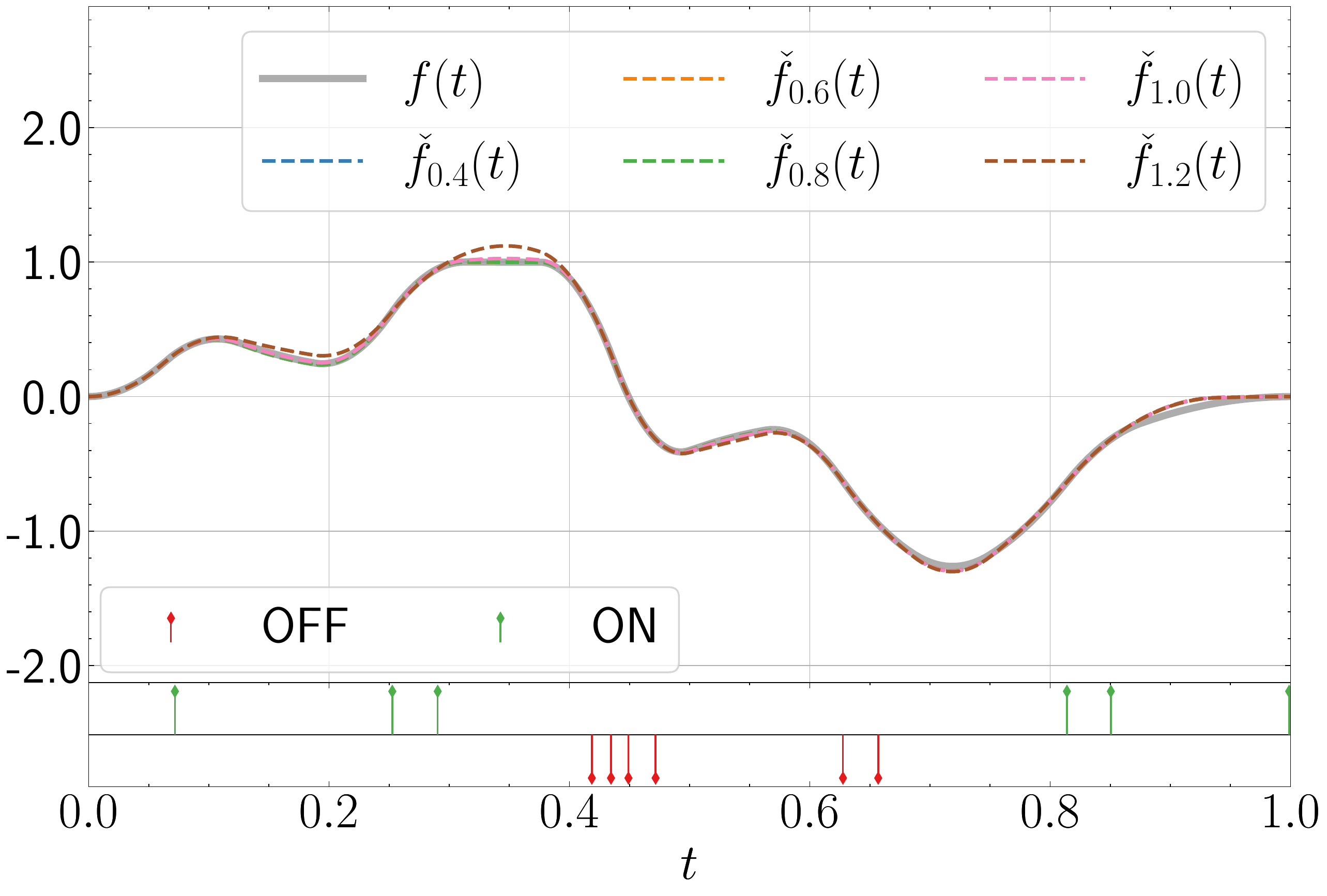}}
	\subfigure[$n=3$, cubic polynomial splines]{\label{fig:lp_reconstruction_3}\includegraphics[width=2.75in]{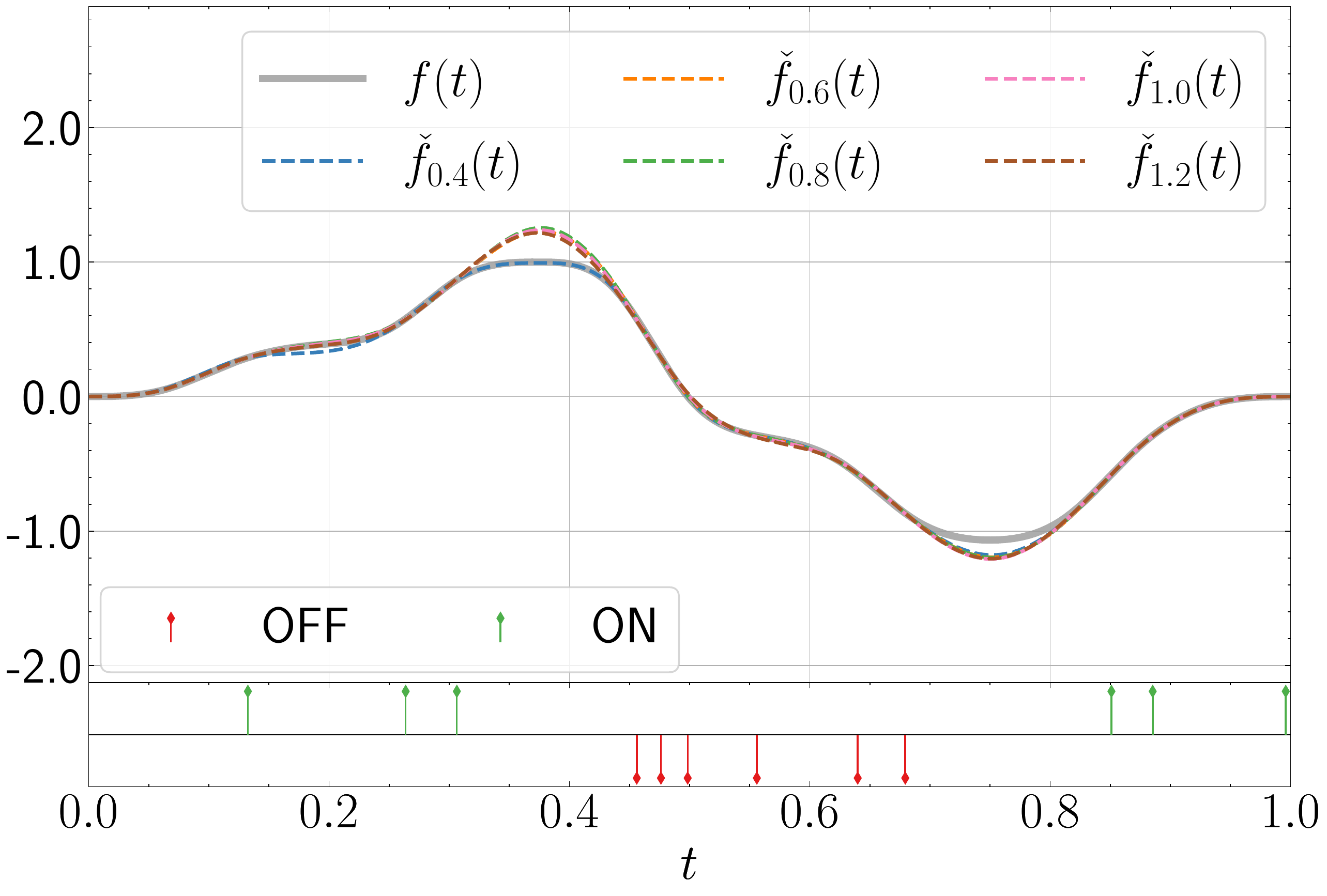}}
	\subfigure[$n=4$, quartic polynomial splines]{\label{fig:lp_reconstruction_4}\includegraphics[width=2.75in]{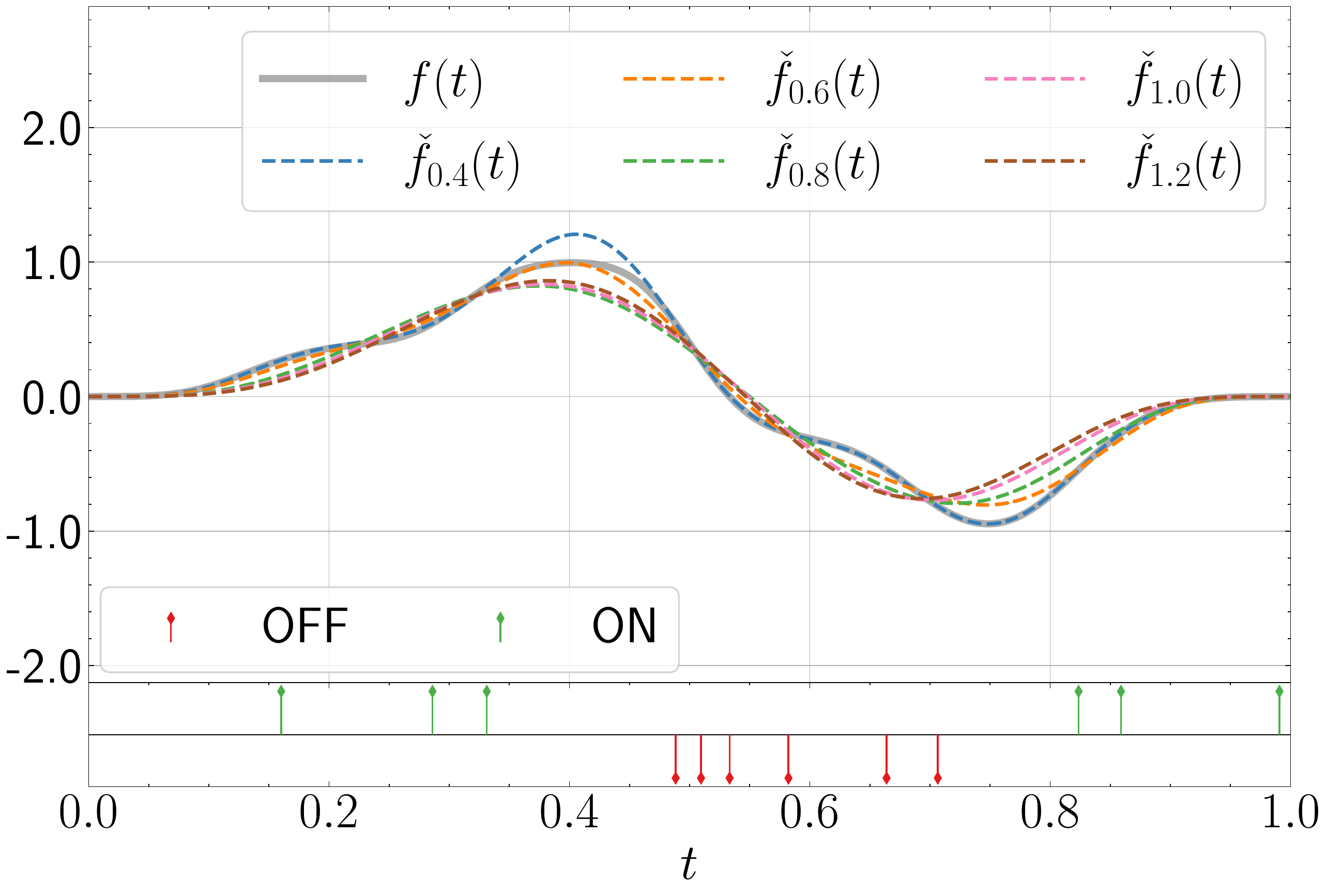}}
	\caption{Signal reconstruction using $L_p$ minimization: the ground truth $f(t)$, and the reconstructions $\check{f}_p(t)$ for $p=0.4,0.6,0.8,1,1.2$ is shown for various choices of $n$: (a) $n=2$, (b) $n=3$, and (c) $n=4$.}
	\label{fig:lp_reconstructions}
\end{figure*}


\subsection{Experimental Results}
\begin{figure}[!t]
	\centering
	\subfigure[]{\label{fig:srer_cumulative}\includegraphics[width=2.75in]{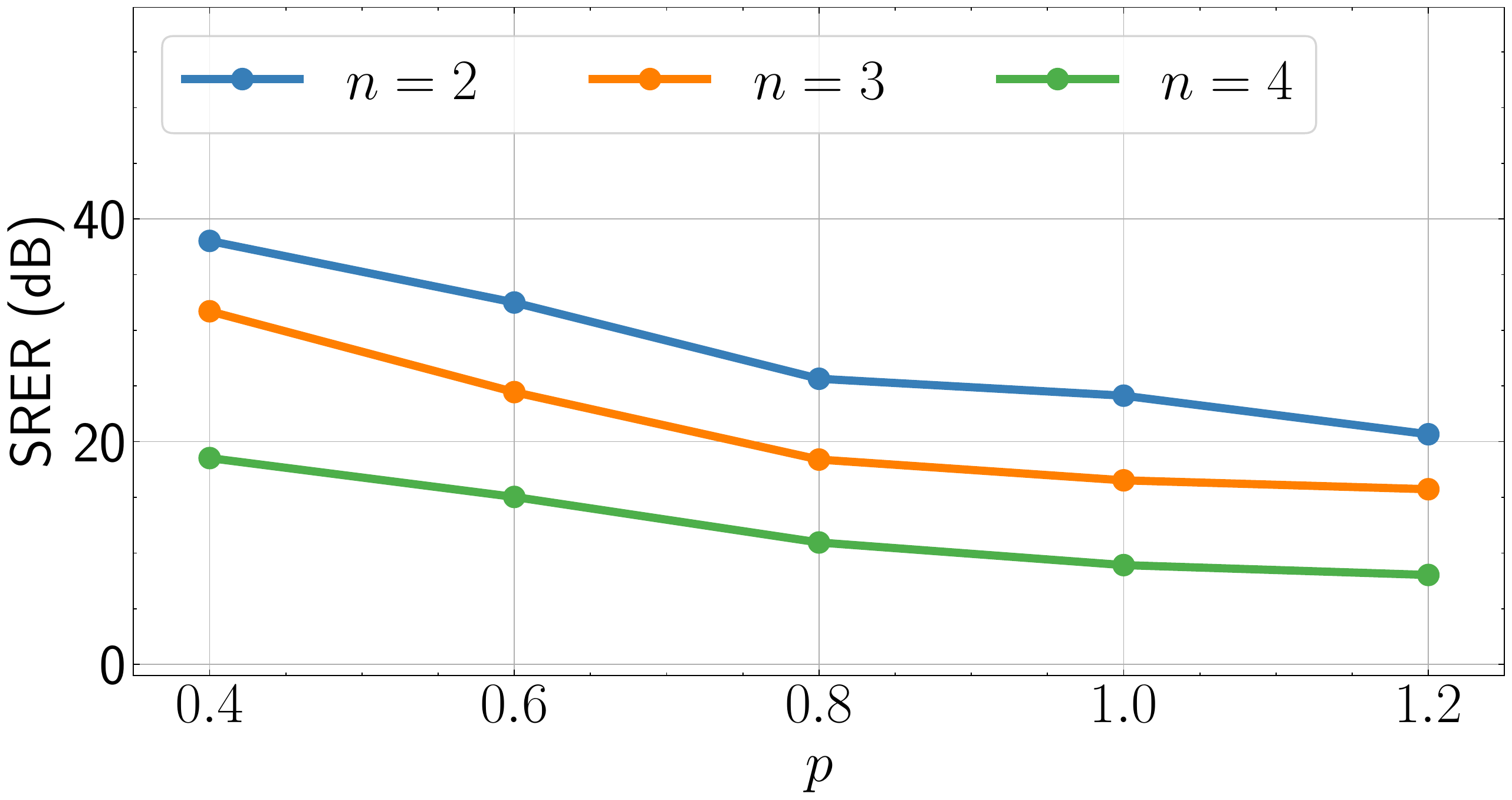}}
	\subfigure[]{\label{fig:overshoot_cumulative}\includegraphics[width=2.75in]{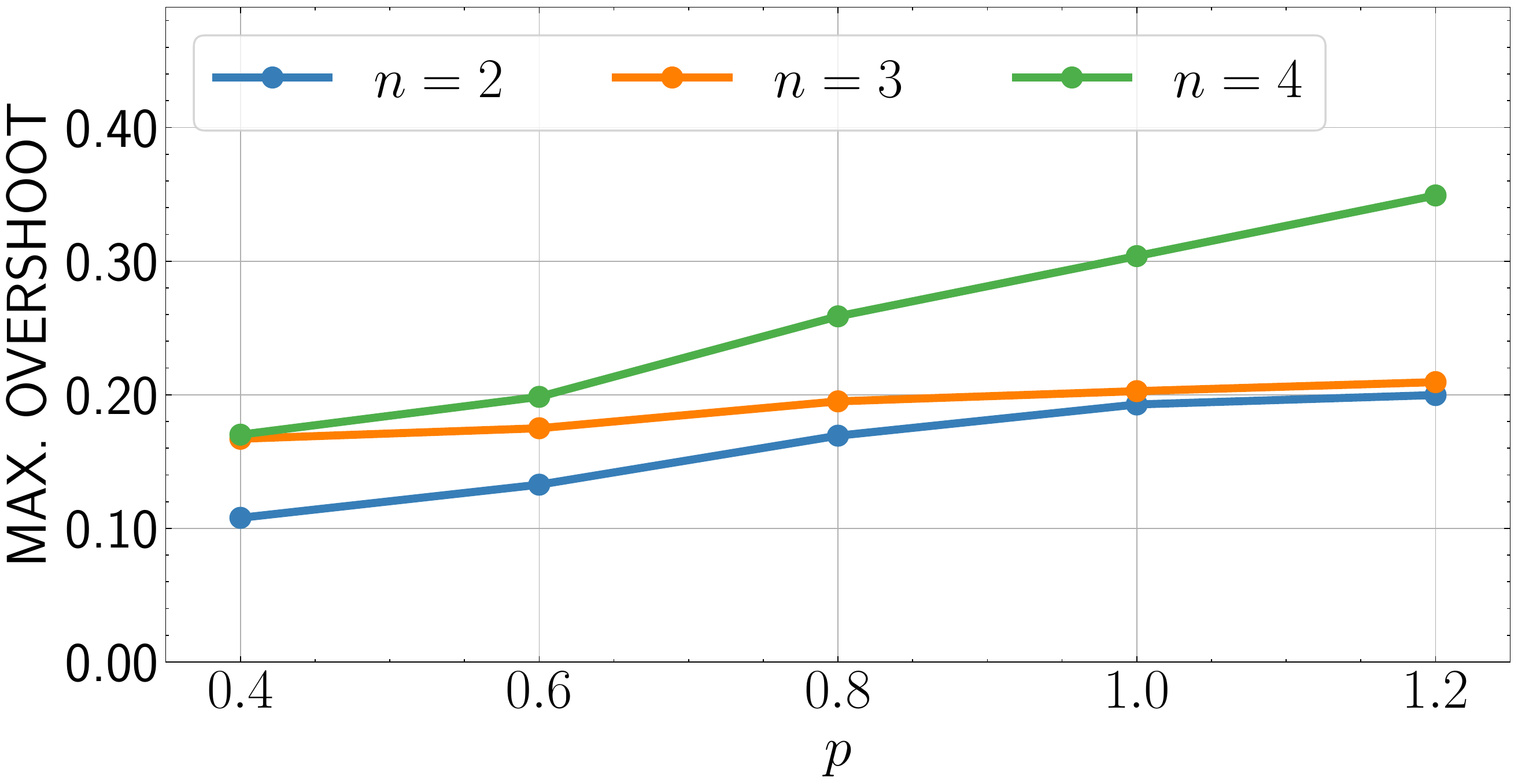}}
	\caption{Objective measures of signal reconstruction: (a) shows the SRER between the ground-truth signal and the reconstruction, and (b) shows the maximum overshoot in the reconstruction, as a function of $p$, and $n=2,3,4$. The results are averaged over $100$ realizations of the ground-truth signal $f$.}
	\label{fig:metrics}
\end{figure}

We consider neuromorphic sampling of signals of the type
\begin{equation*}
	f(t) = \sum_{k=0}^{K-1} c_k \beta_{n,h}(t-hk),
\end{equation*}
in $\cl V_K(\beta_{n,h})$, where $n=2,3,4$, the grid-size $h=2^{-4}$, and support $[0,1]$, with temporal contrast threshold $C$ set to $15\%$ of the dynamic range of the signal. The generated event instants do not meet the sampling density criterion for perfect reconstruction. We perform reconstruction using Algorithm~\ref{algo:reconstruction_using_lp} with $\lambda=10^{-10}$, $\rho=10^{-5}$, and $p=0.4,0.6,0.8,1,1.2$. The stopping criterion is based on how close two consecutive iterates of $\bd c$ get, i.e., the iterations are run until $\norm{\bd c_{k+1}-\bd c_{k}}_2 < 10^{-12}$. In our experiments, we found that this criterion was always met in about 100 iterations. As a computational safeguard, we set an over-riding upper-limit of $500$ on the number of iterations.\\
\indent Figures~\ref{fig:lp_reconstruction_2},~\ref{fig:lp_reconstruction_3}~and~\ref{fig:lp_reconstruction_4} show the ground truth $f(t)$, the reconstructions $\check{f}_p(t)$ for the values of $p$ selected, and the neuromorphic events obtained, for $n = 2, 3,$ and $4$, respectively. The reconstruction quality improves with decreasing $p$, and further, the overshoot in reconstruction in intervals of inactivity in the ground-truth signal is small for small values of $p$. Also, the maximum overshoot in reconstruction decreases with decreasing $p$, which results in a better quality reconstruction. The objective performance measures are signal-to-reconstruction error ratio (SRER) between the ground truth $f$ and the reconstruction $\check{f}$ defined as follows:
\begin{equation}
	\text{SRER} = 10 \log_{10}\left(\frac{\norm{f}^2_{L_2}}{\norm{f-\check{f}}^2_{L_2}}\right) \text{dB},
\end{equation}
and the maximum overshoot between the ground-truth $f$ and the reconstruction $\check{f}$ defined as 
\begin{equation}
\text{Maximum Overshoot} = \norm{f-\check{f}}_{L_\infty}
\end{equation}
Figure~\ref{fig:srer_cumulative} shows the SRER between the ground truth $f(t)$ and the reconstructions $\check{f}_p(t)$ versus $p$, for $n=2,3,4$. The SRER improves with decreasing $p$, for all values of $n$ chosen. The observation corroborates with Figure~\ref{fig:overshoot_cumulative}, which shows that the maximum overshoot in reconstruction decreases with decreasing $p$, thereby improving the reconstruction quality.


\section{Conclusions}
We considered a mathematical model for neuromorphic sampling and made connections with time-encoding by providing a $t$-transform that maps the stream of events to the amplitude samples of the input signal. This brings the problem of signal reconstruction within the framework of nonuniform sampling. We considered neuromorphic sampling of signals in shift-invariant spaces, in particular, signals in Paley-Wiener space and polynomial spline spaces, and analyzed the reconstruction using an alternating-projections approach to show that perfect reconstruction is possible under dense sampling conditions. We provided the necessary and sufficient conditions on the temporal contrast threshold for perfect reconstruction. Next, we considered the scenario where the necessary conditions on the temporal contrast threshold are not met, which may be the case in practice. In this scenario, we posed the signal reconstruction problem as a continuous-domain $L_p$ minimization problem with $p>0$, restricted to the compactly supported class of polynomial spline signals, where the continuous-domain problem can be solved using an equivalent finite-dimensional optimization problem using a variable-splitting approach. Experimental validations on synthetic signals show that the reconstruction accuracy in terms of the SRER and maximum overshoot improve when $p$ decreases.


%
%


\appendices

\section{Proof of Theorem~\ref{thm:general_theorem}}
\label{appendix:proof_general_theorem}
Sandberg's theorem applies to generic continuous-domain linear inverse problems, and to signals in Hilbert spaces, and the proof is available in \cite{sandberg1963properties}. We show a simpler proof that follows \cite{gontier2014sampling}, for signals in $h$-shift-invariant spaces $\cl V(\varphi)$. We begin by recalling the following results.
\begin{lemma}[Wirtinger's inequality]\label{lem:wirtinger}
Let $f, \mathrm{D}\{f\}\in L_2([a,b])$ with $f(a)=0$ or $f(b)=0$. Then,
$$
	\int_a^b |f(u)|^2 \dd u \leq \frac{4}{\pi^2} (a-b)^2 \int_a^b |\mathrm{D}\{f\}(u)|^2 \dd u.
$$
\end{lemma}
\begin{lemma}[Chen, Han, and Jia \cite{chen2005simple}]\label{lem:chen}
Let $\varphi_h,\mathrm{D}\{\varphi_h\}\in L_2(\bb R)$. Then,
$$
	\norm{\mathrm{D}\{f\}}_{L_2} \leq \sup_{\omega\in[-\frac{\pi}{h},+\frac{\pi}{h}]}\frac{G_{\mathrm{D}\{\varphi_h\}}(\omega)}{G_{\varphi_h}(\omega)} \norm{f}_{L_2}.
$$
\end{lemma}
The square of the $L_2$-error in the piecewise-constant approximation is given as
\begin{equation*}
\begin{split}
	\norm{f-\Gamma\{f\}}_{L_2}^2 &= \int_{-\infty}^{+\infty} | f(t) - \Gamma\{f\}(t) |^2 \;\dd t,\\
	&= \sum_{m\in\bb N} \int_{s_m}^{s_{m+1}} |f(t) - f(t_m)|^2 \;\dd t,\\
	= \sum_{m\in\bb N}\!\int_{s_m}^{t_{m}}\!|f(t)&\!-\!f(t_m)|^2\dd t\!+\!\sum_{m\in\bb N}\!\int_{t_m}^{s_{m\!+\!1}}\!|f(t)\!-\!f(t_m)|^2\dd t, \\
	&\overset{(a)}{\leq} \frac{\mathfrak{D}(\{t_m\}_{m\in\bb N})^2}{\pi^2} \norm{\mathrm{D}\{f\}}_{L_2}^2, \\
	&\overset{(b)}{\leq} \gamma^2 \norm{f}_{L_2}^2,
\end{split}
\end{equation*}
where $(a)$ follows from Lemma~\ref{lem:wirtinger}, and $(b)$ follows from Lemma~\ref{lem:chen}. The operator norm $\norm{I-\Pi_{\varphi_h}\Gamma}$ is bounded as follows:
\begin{equation*}
\begin{split}
	\norm{f-(\Pi_{\varphi_h}\Gamma)\{f\}}_{L_2} &= \norm{\Pi_{\varphi_h}\{f-\Gamma\{f\}\}}_{L_2}, \\
	&= \norm{f-\Gamma\{f\}}_{L_2} \leq \gamma \norm{f}_{L_2}.
\end{split}
\end{equation*}
Consider
\begin{equation*}
\begin{split}
	f-f_{k+1} &= f-f_{1} - (I-\Pi_{\varphi_h}\Gamma)\{f_{k}\}, \\
	&= (I-\Pi_{\varphi_h}\Gamma)\{f\} + (I-\Pi_{\varphi_h}\Gamma)\{f_{k}\}, \\
	&= (I-\Pi_{\varphi_h}\Gamma)\{f-f_{k}\} = (I-\Pi_{\varphi_h}\Gamma)^{k+1}\{f\}.
\end{split}
\end{equation*}
\begin{equation*}
\begin{split}
\text{Therefore,}\quad\norm{f-f_{k}}_{L_2} &\leq \norm{I-\Pi_{\varphi_h}\Gamma}^k \norm{f}_{L_2}\leq \gamma^k\norm{f}_{L_2},
\end{split}
\end{equation*}
i.e., $f_{k} \rightarrow f$ in $L_2$-norm, as $k\rightarrow \infty$ when $\gamma<1$.
$\hfill\blacksquare$


\section{Proof of Corollary~\ref{cor:bandlimited}}
\label{appendix:proof_cor_bandlimited}
We begin by showing that the derivative of $f\in\cl V(\sinc_h)$ is bounded.
\begin{lemma}\label{lem:bounded_derivative}
Let $f\in\cl V(\sinc_h)$. Then,
$$
	|\mathrm{D}\{f\}(t)| \leq \frac{\pi}{h^{3/2}} \norm{f}_{L_2}, \; \forall t\in\bb R.
$$
\begin{proof}
Let $f\in\cl V(\sinc_h)$. We have
\begin{equation*}
\begin{split}
	|\mathrm{D}&\{f\}(t)| = \abs{\frac{1}{2\pi}\int_{-\infty}^\infty \jj\omega \hat{f}(\omega)\dd\omega}\leq \frac{1}{2\pi} \int_{-\frac{\pi}{h}}^{+\frac{\pi}{h}} \abs{\omega} \abs{\hat{f}(\omega)}\dd\omega,\\
	&\leq \frac{1}{2\pi} \frac{\pi}{h} \int_{-\frac{\pi}{h}}^{+\frac{\pi}{h}} \abs{\hat{f}(\omega)} \;\dd\omega,\\
	&\overset{(a)}{\leq} \frac{1}{2h} \left(\int_{-\frac{\pi}{h}}^{+\frac{\pi}{h}} \abs{\hat{f}(\omega)}^{2} \dd\omega \right)^{\frac{1}{2}} \left(\int_{-\frac{\pi}{h}}^{+\frac{\pi}{h}} \abs{\mathbbm{1}_{[-\frac{\pi}{h},\frac{\pi}{h}]}(\omega)}^2\dd\omega\right)^{\frac{1}{2}},\\
	&= \frac{1}{2h}\norm{\hat{f}}_{L_2} \left( \frac{2\pi}{h}\right)^{\frac{1}{2}}
	\overset{(b)}{=} \frac{\pi}{h^{3/2}}\norm{f}_{L_2},
\end{split}
\end{equation*}
where $(a)$ follows from Cauchy-Schwartz inequality, and $(b)$ follows from Parseval's identity.
\end{proof}
\end{lemma}
Using Cauchy's mean-value theorem \cite{rudin1953principles}, $\exists \tau\in ]t_{m+1},t_m[$ such that
\begin{equation*}
\begin{split}
	\abs{f(t_{m+1})-f(t_m)} &\leq \abs{\mathrm{D}\{f\}(\tau)}\abs{t_{m+1}-t_m}, \\
	&\leq \frac{\pi}{h^{3/2}}\norm{f}_{L_2} \abs{t_{m+1}-t_m}, \;\forall m\in\bb N.
\end{split}
\end{equation*}
Suppose the sequence $\{f_{k}\}_{k\in\bb N}$ converges to $f$, using Theorem~\ref{thm:general_theorem}, we have
\begin{equation*}
\begin{split}
	&\gamma = \left(\sup_{\omega\in[-\frac{\pi}{h},+\frac{\pi}{h}]}\frac{G_{\mathrm{D}\{\varphi_h\}}(\omega)}{G_{\varphi_h}(\omega)}\right) \frac{\mathfrak{D}(\{t_m\}_{m\in\bb N})}{\pi} <  1,
\end{split}
\end{equation*}
which implies that $\mathfrak{D}(\{t_m\}_{m\in\bb N}) < h$. Then, we have
\begin{equation*}
\begin{split}
	h > \mathfrak{D}(\{t_m\}_{m\in\bb N}) &\geq \abs{t_{m+1}-t_m},\\
	&\geq \abs{f(t_{m+1})-f(t_m)}\frac{h^{\frac{3}{2}}}{\pi\norm{f}_{L_2}}.
\end{split}
\end{equation*}
Using Definition~\ref{def:neuromorphic_encoder}, we obtain $\displaystyle h>\abs{p_m C}\frac{h^{\frac{3}{2}}}{\pi\norm{f}_{L_2}}$, which implies $C < \frac{\pi}{\sqrt{h}} \norm{f}_{L_2}$ as $p_m\in\{-1,+1\}$.
$\hfill\blacksquare$

\section{Proof of Corollary~\ref{cor:spline}}
\label{appendix:proof_cor_spline}
We show that the derivative of $f\in\cl V(\beta_{n,h})$ is bounded.
\begin{lemma}
Let $f = \mathrm{V}_{\beta_{n,h}}\bld c\in\cl V(\beta_{n,h})$ and $n\geq 1$. Then,
\begin{equation*}
	|\mathrm{D}\{f\}(t)| \leq \frac{2}{h} \norm{\bld c}_{\ell_1}, \; \forall t\in\bb R.
\end{equation*}
\begin{proof}
Consider $f = \mathrm{V}_{\beta_{n,h}}\bld c$. Then,
\begin{equation*}
\begin{split}
\abs{\mathrm{D}\{f\}(t)} &= \abs{\sum_{k\in\bb Z}c_k \frac{1}{h}\!\left(\beta_{n\!-\!1,h}(t\!-\!kh)\!-\!\beta_{n\!-\!1,h}(t\!-\!(k\!+\!1)h)\right)},\\
&\leq \frac{1}{h} \sum_{k\in\bb Z} \abs{c_k}\!\left( \abs{\beta_{n\!-\!1,h}(t\!-\!kh)} \!+\! \abs{\beta_{n\!-\!1,h}(t\!-\!kh\!-\!h)} \right),\\
&\overset{(a)}{\leq} \frac{2}{h} \sum_{k\in\bb Z} \abs{c_k} = \frac{2}{h} \norm{\bld c}_{\ell_1},
\end{split}
\end{equation*}
where $(a)$ follows using $\abs{\beta_{n,h}(t)}\leq 1, \forall t\in\bb R, n\in\bb N$.
\end{proof}
\end{lemma}
Next, we compute the constant $\gamma$ for signals in $\cl V(\beta_{n,h})$. The Fourier transform of ${\beta}_{n,h}(t)$ is given by 
\begin{equation*}
	\hat{\beta}_{n,h}(\omega) = h \left( \frac{1-e^{-\jj\omega h}}{\omega h} \right)^{n+1}.
\end{equation*}
Therefore, we have the quotient function
\begin{equation*}
\begin{split}
	\frac{G^2_{\mathrm{D}\{\beta_{n,h}\}}(\omega)}{G^2_{\beta_{n,h}}(\omega)} &= \frac{\displaystyle\sum_{k\in\bb Z} \abs{\jj\left(\omega + \frac{2\pi}{h}k\right)\hat{\beta}_{n,h}\left(\omega + \frac{2\pi}{h}k\right)}^2}{\displaystyle\sum_{k\in\bb Z} \abs{\hat{\beta}_{n,h}\left(\omega + \frac{2\pi}{h}k\right)}^2}, \\
	&= \frac{\displaystyle\sum_{k\in\bb Z}\abs{\frac{1}{\omega h+2\pi k}}^{2n}}{h^2\displaystyle\sum_{k\in\bb Z}\abs{\frac{1}{\omega h+2\pi k}}^{2n+2}},
\end{split}
\end{equation*}
which is $\frac{\pi}{h}$-periodic, and attains its maximum value at $\omega=\frac{\pi}{h}$. Therefore, we have
\begin{equation*}
\begin{split}
	\sup_{\omega\in[-\frac{\pi}{h},+\frac{\pi}{h}]}\frac{G^2_{\mathrm{D}\{\beta_{n,h}\}}(\omega)}{G^2_{\beta_{n,h}}(\omega)} &= \frac{\pi^2\displaystyle\sum_{k\in\bb Z}\abs{\frac{1}{2k+1}}^{2n}}{h^2\displaystyle\sum_{k\in\bb Z}\abs{\frac{1}{2k+1}}^{2n+2}},\\
	&= \frac{4\pi^2\left(2^{2n}-1\right)\zeta(2n)}{h^2\left(2^{2n+2}-1\right)\zeta(2n+2)},
\end{split}
\end{equation*}
where $\displaystyle\zeta(x) = \sum_{k=1}^\infty\frac{1}{k^x}$ denotes the Riemann zeta function. Therefore, following Theorem~\ref{thm:general_theorem} for perfect reconstruction, the sampling density must be bounded above as follows:
\begin{equation*}
	\mathfrak{D}(\{t_m\}_{m\in\bb N}) \leq \frac{h}{2} \underbrace{\sqrt{\frac{\left(2^{2n+2}-1\right)\zeta(2n+2)}{\left(2^{2n}-1\right)\zeta(2n)}}}_{\eta}.
\end{equation*}
Using Cauchy's mean-value theorem \cite{rudin1953principles}, $\exists\tau\in\quad]t_{m+1},t_m[$ such that
\begin{equation*}
\begin{split}
	\abs{f(t_{m+1})-f(t_m)} &\leq \abs{\mathrm{D}\{f\}(\tau)}\abs{t_{m+1}-t_m}, \\
	&\leq \frac{2}{h} \norm{\bld c}_{\ell_1}\abs{t_{m+1}-t_m}, \;\forall m\in\bb N.
\end{split}
\end{equation*}
Then, we have
\begin{equation*}
\begin{split}
	h\frac{\eta}{2} > \mathfrak{D}(\{t_m\})\!\geq\!\abs{t_{m+1}\!-\!t_m}\!&\geq\!\abs{f(t_{m+1})\!-\!f(t_m)}\frac{h}{2\norm{\bld c}_{\ell_1}}, \\
	&\overset{(a)}{=} \abs{p_m C}\frac{h}{2\norm{\bld c}_{\ell_1}}, \\
\end{split}
\end{equation*}
where $(a)$ follows from Definition~\ref{def:neuromorphic_encoder}, and thus $C < \eta \norm{\bld c}_{\ell_1}$ as $p_m\in\{-1,+1\}$. Further, suppose that $f$ has a finite support with $K$ expansion coefficients, then, using the Riesz basis condition given in Eq.~\eqref{eq:riesz_bounds}, we have
\begin{equation*}
\norm{\bld c}_{\ell_1} \leq \sqrt{K} \norm{\bld c}_{\ell_2} \leq \sqrt{K}\left(\frac{\pi}{2h}\right)^{n+1} \norm{f}_{L_2}.
\end{equation*}
Correspondingly, we obtain $\displaystyle C < \sqrt{K}\left(\frac{\pi}{2h}\right)^{n+1}\eta\norm{f}_{L_2}$.
$\hfill\blacksquare$

\section{Proof of Lemma~\ref{lem:finite_updates}}
\label{appendix:proof_finite_updates}
Consider $g(t) = (\Pi_{\varphi_h}\Gamma)\{f\}(t)$, where $f = \mathrm{V}_{\varphi_h}\bld a$. Using Eq.~\ref{eq:projection_operator}, we write
\begin{align*}
	g(t) &= \sum_{k\in\bb Z}\sum_{j\in\bb N}f(t_j)\langle \mathbbm{1}_{[s_j,s_{j+1}]},\tilde{\varphi}_h(\cdot-hk)\rangle \varphi_h(t-hk),\\
	&= \sum_{k\in\bb Z} b_k\varphi_h(t-hk),
\end{align*}
where $b_k = \sum_{j\in\bb N} f(t_j)\int_{s_j}^{s_{j+1}} \tilde{\varphi}_h(u-hk)\dd u$,
\begin{align*}
	&= \sum_{j\in\bb N}\left(\sum_{k\in\bb Z} a_k \varphi_h(t_j-hk)\right) \int_{s_j}^{s_{j+1}} \tilde{\varphi}_h(u-hk)\dd u,\\
	&= \sum_{j\in\bb N} (H\bld a)_j \tilde{H}_{k,j},
\end{align*}
or, succinctly expressed as $\bld b = \tilde{H}H\bld c$. $\hfill\blacksquare$


\bibliographystyle{IEEEbib}
\bibliography{references.bib}

\end{document}